\documentclass[twocolumn,showpacs,prb,eqsecnum,citeautoscript,amsmath,amssymb,floatfix,superscriptaddress]{revtex4-1}
\usepackage{bm,color,amsmath,amssymb,mathrsfs,latexsym,graphicx,psfrag,float}

\newcommand{\opt}[1]

\usepackage{comment}
\usepackage{hyperref}
\usepackage{enumerate}
\usepackage{txfonts}
\usepackage{bbm}
\usepackage{ifthen}
\usepackage{cleveref}
\usepackage{dsfont}
\usepackage{tabularx}
\newcolumntype{L}[1]{>{\raggedright\arraybackslash}p{#1}} % linksbündig mit Breitenangabe
\newcolumntype{C}[1]{>{\centering\arraybackslash}p{#1}} % zentriert mit Breitenangabe
\newcolumntype{R}[1]{>{\raggedleft\arraybackslash}p{#1}} % rechtsbündig mit Breitenangabe

\newcommand{\eq}[2]{\begin{align}\label{#1} #2 \end{align}}

\newcommand{\cre}[2]{\bar{#1}_{#2}}

\newcommand{\fie}[2]{#1_{#2}}
\newcommand{\cfie}[2]{\bar{#1}_{#2}}
\newcommand{\ann}[2]{#1_{#2}}
\newcommand{\sub}[1]{_{\mbox{\tiny #1}}}

\newcommand{\red}[1]{\textcolor{black}{#1}}

\usepackage{color}

\begin{document}

\title{Kinetic Theory for Interacting Luttinger Liquids}

\author{Michael Buchhold}
\affiliation{Institut f\"ur Theoretische Physik, Technische Universit\"at Dresden, 01062 Dresden, Germany}
\affiliation{Institut f\"ur Theoretische Physik, Leopold-Franzens Universit\"at Innsbruck, A-6020 Innsbruck, Austria}
\author{Sebastian Diehl}
\affiliation{Institut f\"ur Theoretische Physik, Technische Universit\"at Dresden, 01062 Dresden, Germany}
\affiliation{Institut f\"ur Theoretische Physik, Leopold-Franzens Universit\"at Innsbruck, A-6020 Innsbruck, Austria}

\begin{abstract}
We derive a closed set of equations for the kinetics and non-equilibrium dynamics of interacting Luttinger Liquids with cubic resonant interactions. In the presence of these interactions, the Luttinger phonons become dressed but still well defined quasi-particles, characterized by a life-time much larger then the inverse energy. This enables the separation of forward time dynamics and relative time dynamics into slow and fast dynamics and justifies the so-called Wigner approximation, which can be seen as a "local-time approximation" for the relative dynamics. Applying field theoretical methods in the Keldysh framework, i.e. kinetic and Dyson-Schwinger equations, we derive a closed set of dynamic equations, describing the kinetics of normal and anomalous phonon densities, the phonon self-energy and vertex corrections for an arbitrary non-equilibrium initial state. In the limit of low phonon densities, the results from self-consistent Born approximation are recaptured, including Andreev's scaling solution for the quasi-particle life-time in a thermal state. As an application, we compute the relaxation of an excited state to its thermal equilibrium. \red{ While the intermediate time dynamics displays exponentially fast relaxation, the last stages of thermalization are governed by algebraic laws. This can be traced back to the importance of energy and momentum conservation at the longest times, which gives rise to dynamical slow modes. }

\end{abstract}


\maketitle
\section{Introduction}
The kinetics and non-equilibrium dynamics of low dimensional, interacting quantum systems is an outstanding and fascinating challenge in quantum many-body physics \cite{cardycalabrese06,esslinger04}. On the one hand, it is strongly motivated by recent cold atom experiments performed on low entropy quantum wires under out-of-equilibrium conditions \cite{kinoshita,schmiedmayer12,schmiedmayernphys12,schmiedmayernjp13,bloch13,nagerl13,trupke13}. On the other hand, from a theoretical point of view, the study of non-equilibrium dynamics in integrable and nonintegrable systems is currently a field of growing interest \cite{cazalilla06,burkov07,caux13}. This is triggered by the question, whether and -- if answered affirmatively -- in which specific way a one-dimensional quantum system is able to thermalize \cite{berges_pretherm}.

An example of a one-dimensional integrable model is the linear Luttinger Liquid, which is the effective long-wavelength description of one-dimensional interacting quantum fluids, composed either of fermions or bosons \cite{haldane81,giamarchi04}. Due to integrability, even if prepared in a non-equilibrium state, this model will never thermalize, since the number of excitations for each momentum mode $q$ is a constant of motion \cite{cazalilla06}. However, as already pointed out by Andreev and Haldane \cite{andreev80,haldane81}, there are non-zero corrections to the linear theory, which certainly break integrability in the Luttinger model. They are irrelevant in the sense of the renormalization group, and do not affect static observables. In contrast, they lead to a modification of {\it dynamical}, i.e. frequency resolved, correlation functions. These nonlinear corrections describe three-body scattering processes between phonons and are, due to their resonant nature, not straightforwardly approached  theoretically. 

Apart from a wealth of numerical studies \cite{caux06,affleck06,pereira08,essler10,wouters14,caibarthel13,schachen14,poletti12}, based on matrix product state and Bethe ansatz calculations, several field theoretical approaches have been developed \cite{samokhin98,narayan02,pustilnik06,pustilnik07,rozhkov05,imambekov08,Matveev13,Matveev14}. A seminal early study was carried out by Andreev, who used a self-consistent Born approximation to determine the phonon self-energies, establishing a universal phonon absorbtion rate $\gamma_q\sim q^{\eta}$ with exponent $\eta=\frac{3}{2}$ for a finite temperature system \cite{lamacraft13,vanBeijeren,gangardt13}. A similar computation leads to an exponent $\eta=2$ for the case of a zero temperature state \cite{zwerger06}, which has been verified by several numerical methods \cite{affleck06}. Recently, so-called nonlinear Luttinger Liquids have been introduced, which are designed to capture corrections to the linear Luttinger theory in the context of one dimensional fermions systems \cite{imambekov09,imambekov12,pustilnik06,pustilnik07}.  These have been very sucessful in determining, for example, the power law divergences of the dynamic structure factor, or thermalization rates for near equilibrium systems \cite{imambekov09}.
Despite the large number of analytic and semi-analytic works \cite{Kennes13,Karrasch12,mitragiamarchi11,Proto14a,cazalilla06,cazalillareview}, only few approaches for far from equilibrium dynamics have been developed so far \cite{mitragiamarchi12,Proto14b,Tavora13}.

The purpose of this article is to provide a quantitative description of the kinetics of an interacting Luttinger Liquid. 
Strong motivation comes from a recent surge of experiments with interacting one-dimensional bosons and fermions in ultracold atom setups performed under out-of-equilibrium conditions \cite{schmiedmayernphys12,schmiedmayer12,trupke13,nagerl13,hild14}. Here a one-dimensional quantum fluid is prepared in a true non-equilibrium state, and the experiment subsequently witnesses the time evolution of the system. At the present stage, the systems' properties are still well described by linear Luttinger Liquid theory with, however, a time dependent, far from equilibrium distribution function \cite{schmiedmayernphys12,schmiedmayer12,trupke13,schmiedmayernjp13}. So far, there are numerous theoretical works which describe these systems in terms of non-interacting Luttinger liquids alone, where there is only dephasing dynamics and the evolution of the distribution function is absent \cite{cazalilla06,Kennes13,Langen14,agarwal14}. This describes well the pre-thermalized regime in the shorter time dynamics. However, current experiments are steadily pushed to larger observation times, where the thermalization crossover caused by the residual (RG irrelevant) interactions in the Luttinger liquid should occur.  What is therefore lacking at present in the theoretical literature, is a kinetic theory for the time evolution of the distribution function in these systems, which is able to track such a crossover and the associated time scales. This is the aim of the present paper.

On the technical side, achieving our goal amounts to obtaining theoretical control over the infrared divergences inherent to naive perturbation theory for the interacting Luttinger Liquid. While this issue was solved for the stationary equilibrium (or near equilibrium, in the sense of linear response) long time ago by Andreev and others \cite{andreev80,samokhin98,zwerger06}, we focus here on getting this problem under control in the kinetic equation, initialized with a general Gaussian non-equilibrium distribution function.

In more detail, exploiting the resonant but subleading character of the interactions in a one-dimensional interacting quantum fluid, we apply non-equilibrium diagrammatic theory to solve for the non-equilibrium dynamics of an interacting Luttinger Liquid. We show that due to the resonant but subleading nature of the interactions, vertex corrections are moderate for many physical realizations and consequently the non-equilibrium dynamics and self-energy can be solved within self-consistent Born approximation. The self-consistency is however, crucial, and a perturbative Born approach leads to infrared divergencies.
The result is an effectively {\it closed} set of equations for the time-dependent phonon density and self-energy in the presence of resonant interactions. This approach, without considering the vertex correction and restricted to equilibrium systems, has been discussed also for non-interacting one-dimensional dispersive fermions \cite{aristov} and in the context of the Coulomb drag effect for sufficiently low temperatures \cite{pustilnik03}.

As a major result of the RG irrelevant but resonant interactions, the excitations remain well-defined but dressed phonons, with a life-time $\tau_q$ much larger than their typical coherent time-scale $\epsilon_q^{-1}$. The dressed spectral function remains sharply peaked at the bare phonon energies $\epsilon_q$ with width $\tau_q^{-1}$, such that the self-energy and distribution function of the phonons for frequencies sufficiently close to the on-shell frequency can be approximated by their on-shell value.
This is referred to as the quasi-particle approximation. 
The long life-time of the dressed phonons results in a further simplification, as it implies that the forward time evolution of the system is much slower then the relative time evolution. This decoupling leads to a "local time approximation", where an effectively stationary problem can be solved at each instant of time. 
 We will quantify the validity of these approximations in terms of general but state dependent quantitative bounds below.

Our estimate of the vertex correction further supports the validity of the self-consistent Born approximation for the time evolution. More precisely, for zero temperature states, the vertex correction vanishes identically, reproducing previous results \cite{forsternelson76}. Away from the ground state, the loop correction leads to a finite multiplicative renormalization of the vertex, which remains small for states close to thermal equilibrium. This implies that for typical translation invariant low entropy initial states, the equations governing the time evolution of the phonon occupation and the self-energy are effectively closed. It does not rule out, however, the possibility of significant vertex renormalization in general.

The strength of this approach is the simplicity of the resulting final equations, which can directly be implemented and solved numerically. This provides a useful tool with a broad spectrum of applicability, ranging from tracking the thermalization process of non-integrable, weakly interacting Bose and Fermi gases to the study of interacting open system dynamics \cite{eckstein09,schachen14,poletti12,poletti13}. It is suited for an initial state of a general Gaussian form with arbitrary, non-zero phonon densities, including diagonal and off-diagonal occupations. 

The structure of the kinetic equation and the equations for the self-energy and vertex correction reveal strong aspects of universal behavior. This concerns two key points: First, for an occupation function, which decreases sufficiently strongly in momentum space, they are independent of the short distance cutoff of the Luttinger liquid, and thus of microscopic details inherited from even shorter length scales. This property emerges via a ``bootstrap'' mechanism within the self-consistent Born approximation, as we elaborate on in the main text. Second, after a proper rescaling of time\footnote{According to $t\rightarrow \frac{t}{v_0}$, where $v_0$ is the strength of the nonlinearity.}, all microscopic parameters are completely eliminated from the dynamic equations. Consequently and remarkably, the only microscopic information that enters the dynamics is the initial phonon density at $t=0$.

At this point, we also mention three limitations of our approach, which however are not directly physically relevant for the ultracold atomic systems we are targeting. First, as already briefly mentioned above, our setting is restricted to initial quantum states with Gaussian, although in general far-from-equilibrium correlations. 
Such conditions have been discussed extensively in previous works\cite{cazalilla06,mitragiamarchi12,Kennes13,schmiedmayernphys12,schmiedmayer12,trupke13},  mostly for the case of linear Luttinger Liquids generating the subsequent dynamics. Second, our method is not suited to describe the asymptotic infrared behavior of a one-dimensional quantum fluid, where for the smallest momenta\footnote{For momenta $k\ll \sqrt{K}\Lambda$, where $K$ is the Luttinger parameter and $\Lambda$ the Luttinger cutoff.} it has been shown that the elementary excitations are fermions\cite{pustilnik06,pustilnik07,Matveev13,Matveev13a,Matveev14} with a strongly suppressed decay rate $\gamma\sim k^8$. Third, it doesn't work for frequencies (or temperatures) above the Luttinger cutoff, where the dominant interaction is given by scattering between Luttinger phonons and mobile electronic impurities\cite{imambekov08,imambekov09a,imambekov09,imambekov12}. However, the range of validity of our approach coincides perfectly with cold atom experiments, which consider momenta $10^{-2}\Lambda < p <\Lambda$ and temperatures $k\sub{B}T<\hbar u\Lambda$, where $u$ is the sound velocity and $\Lambda$ the Luttinger cutoff. The Luttinger cutoff scale is given by the chemical potential for these experiments, $\Lambda \approx \sqrt{2 g n_0}$ for weak interactions $g$, and $n_0$ the mean density. The infrared restriction of the momentum range results from typical trap sizes, with oscillator lengths  roughly two orders of magnitude larger than $1/\Lambda$. This discards precision measurements of frequency resolved observables in these extreme infrared asymptotic regimes, and thus is not of foremost interest for our study. 

The remainder of the paper is organized as follows. In Sec.~\ref{sec:Sec2}, we introduce and briefly discuss the action of the interacting Luttinger model in the phonon representation and in the Keldysh non-equilibrium framework. In Sec.~\ref{sec:Sec3}, we derive the non-equilibrium fluctuation-dissipation relation (FDR) and determine the phonon self-energies in self-consistent Born approximation for an arbitrary phonon distribution. Furthermore, we discuss the necessary approximations and quantify their justification.
Subsequently, taking advantage of Sec.~\ref{sec:Sec3}, we determine the kinetic equation for the phonon density in self-consistent Born approximation in Sec.~\ref{sec:Sec4}.
In Sec.~\ref{sec:Sec5}, we take into account non-zero off-diagonal (anomalous) phonon densities and show in which way the kinetic equation and the phonon self-energy are modified in their presence. Additionally, we derive a kinetic equation for the anomalous phonon densities.
These results are applied in Sec.~\ref{sec:Sec6} to determine the relaxation of an excited, thermal state back to thermal equilibrium. Furthermore, the analytically obtained relaxation rate is compared to the numerical value, showing excellent agreement.
Finally, in Sec.~\ref{sec:DS}, we go beyond the self-consistent Born approximation and apply Dyson-Schwinger equations to take into account a non-zero vertex correction. We determine a closed set of equations for the kinetic equation, self-energy and vertex correction for arbitrary states, and discuss the effect of the latter.

\section{Model}\label{sec:Sec2}
The action describing the interacting Luttinger model consists of two parts (we set $\hbar=1$)
\eq{Model1}{S=S\sub{TL}+S\sub{Int}.
}
Here, $S\sub{TL}$ is the well-known quadratic Tomonaga-Luttinger (TL) action \cite{tomonaga50,luttinger63,haldane81}
\eq{Model2}{
S\sub{TL}=\frac{1}{2\pi}\int_{x,t} \left[\left(\partial_x \phi\right)\left(\partial_t\theta\right)-u  K\left(\partial_x\theta\right)^2-\frac{u}{K}\left(\partial_x\phi\right)^2\right],
}
where $\int_{x,t}\equiv \int_{-\infty}^{\infty}dt\ dx$ is the integral over space and time and $\phi=\phi(x,t)$ and $\theta=\theta(x,t)$ are dimensionless, real fields. 
The non-linear part $S\sub{Int}$ is cubic in the fields and reads \cite{haldane81b}
\eq{Model3}{
S\sub{Int}=\frac{1}{2\pi}\int_{x,t}\left[ \kappa\sub{bc}\left(\partial_x\theta\right)^2\left(\partial_x\phi\right)+\kappa\sub{qp}\left(\partial_x\phi\right)^3\right].}
Starting from a microscopic derivation of the TL model as the effective long-wavelength description of interacting bosons or fermions in one dimension, the fields $\theta, \phi$ represent local phase and density fluctuations \cite{haldane81,giamarchi04}. In this setting, we consider effective electron-electron (or boson-boson) interaction to be short ranged, i.e. of $\delta$-function type\cite{giamarchi04}.
 The non-linearity corresponding to $\kappa\sub{bc}$ as well is of microscopic origin and is referred to as band curvature. It originates from deviations from a perfectly linear dispersion of the microscopic particles. On the other hand, the term corresponding to $\kappa\sub{qp}$ is generated in an effective long-wavelength description, where the fast modes have been integrated out already, and describes effective three-particle interactions. 

The fields $\theta, \phi$ are dimensionless, i.e. they have a canonical scaling dimension equal to zero. As a result, they do not scale when coarse graining to larger distances, i.e. when performing the rescaling 
\eq{Model4}{
x\rightarrow l x, \ \ \ t\rightarrow l^z t,}
where $z=1$ is the dynamical exponent and $l>1$. In contrast
\eq{Model5}{
S\sub{Int}\rightarrow \frac{1}{2\pi l}S\sub{Int}}
under the rescaling \eqref{Model4}, such that the influence of $S\sub{Int}$ vanishes on the longest wavelengths, i.e. it becomes irrelevant in the renormalization group (RG) sense. Consequently, the static equilibrium properties of the interacting Luttinger model (Eq.~\eqref{Model1}) are well described by the quadratic part of the action alone and the partition function can be approximated by
\eq{Model6}{
Z=\int\mathcal{D}[\theta,\phi]\ e^{iS}\approx \int\mathcal{D}[\theta,\phi]\ e^{iS\sub{TL}}.}
Here, $\int\mathcal{D}[\theta,\phi]$ stands for the functional integral over the fields $\theta, \phi$.

The Tomonaga-Luttinger action describes phonons with a dispersion $\epsilon_q=u |q|$ linear in the momentum $|q|$, propagating with the speed of sound $u$. In the absence of $S\sub{int}$, these phonons are non-interacting and consequently the phonon density for a specific mode $q$ is a conserved quantity. However, although the phonon interaction is irrelevant in the RG sense, it contains resonant processes where two phonons propagating in the same direction and the same speed of sound can interact with each other for an infinite time span. This leads to a non-trivial modification of time-dependent, dynamical observables compared to the case of non-interacting phonons. As pointed out in a seminal work by Andreev \cite{andreev80} (considering finite $T$) and more recent work\cite{zwerger06,affleck06}, the presence of $S\sub{Int}$ leads to a finite phonon lifetime
\eq{Model7}{
\tau_q\sim |q|^{-\eta}\mbox{ with }\left\{ \begin{array}{lc}\eta=\frac{3}{2}& \mbox{ for }T>0\\ \eta=2& \mbox{ for }T=0\end{array}\right. ,
}
which, in equilibrium, is visible only in dynamical, i.e. frequency dependent quantities such as the dynamical structure factor 
\eq{Model8}{
S(q,\omega)=\int_{x,t} e^{i(\omega t-qx)}\langle \left(\partial_x\phi\right)_{x,t}\left(\partial_x\phi\right)_{0,0}\rangle,}
with 
\eq{Model9}{
\langle ... \rangle=\frac{1}{Z}\int\mathcal{D}[\theta,\phi]\ ... \ e^{iS}.}
For dynamical quantities it is therefore important to take into account the full action, Eq.~\eqref{Model1}, instead of the reduced quadratic part only, as indicated in Eq.~\eqref{Model6}.

For a true non-equilibrium situation, the phonon distribution is not stationary, i.e. not a thermal or zero temperature distribution, but instead the phonon number $n_q$ for a given momentum mode $q$ becomes time-dependent. The redistribution of phonons between the different momentum modes with exact energy conservation is described by the resonant interactions in $S\sub{Int}$, and in a non-equilibrium situation the action can not be reasonably reduced to the quadratic Tomonaga-Luttinger action, for which the phonon density is a constant of motion.

Since we are interested in the non-equilibrium dynamics in the interacting TL model, we formulate the problem in a Keldysh path integral framework, which is able to treat both equilibrium and non-equilibrium dynamics on equal footing \cite{KeldyshPrime,kamenevbook,kamenev09}. We will now shortly introduce the canonical Bogoliubov transformation, which switches from the basis of real fields $\theta, \phi$ to the basis of complex fields $\cre{a}, a$. Those correspond to creation and annihilation operators in an operator picture \cite{giamarchi04}. We close the model section by placing the action \eqref{Model1} on the Keldysh contour and briefly explaining the formalism.

\subsection{Phonon basis}
In order to use a physically more appealing representation of the TL action, one commonly introduces a set of complex fields $\cre{a}, a$ which represent the (bosonic) eigenmodes of the system, i.e. the discussed phonons. The corresponding Bogoliubov transformation is 
\begin{eqnarray}
\theta_{x,t}&=&\theta_0+\frac{i}{2}\int_q \left(\frac{2\pi}{|q|K}\right)^{\frac{1}{2}}e^{-iqx}\left(\cfie{a}{q,t}-\fie{a}{-q,t}\right),\label{Model10}\\
\phi_{x,t}&=&\phi_0-\frac{i}{2}\int_q\left(\frac{2\pi K}{|q|}\right)^{\frac{1}{2}}\mbox{sgn}(q) e^{-iqx}\left(\cfie{a}{q,t}+\fie{a}{-q,t}\right),\ \ \ \ \ \ \label{Model11}
\end{eqnarray}
with  abbreviations $\phi_{x,t}=\phi(x,t)$ and $\int_q=\int_{-\infty}^{\infty}\frac{dq}{2\pi}$ and the Fourier transformed phonon fields
\eq{Model12}{
\fie{a}{q,t}=\int_x e^{-iqx} \fie{a}{x,t}.}
The product $\cfie{a}{x,t}\fie{a}{x,t}$ represents a phonon density and, therefore, in the continuum limit, the fields $\cfie{a}{x,t}, \fie{a}{x,t}$ are not dimensionless, in contrast to $\phi_{x,t}, \theta_{x,t}$, but scale as $\cfie{a}{x,t}\sim \frac{1}{\sqrt{x}}$.
The quadratic part of the action transforms into
\eq{Model13}{
S\sub{TL}=\frac{1}{2\pi}\int_{q,t} \cfie{a}{q,t}\left( i\partial_t-u|q|\right)\fie{a}{q,t},}
describing non-interacting phonons with a linear dispersion $\epsilon_q=u|q|$.
The cubic part becomes
\begin{eqnarray}
S\sub{Int}&=&\frac{1}{2\pi}\int_{q,p,t}v_{q,p,p+q}\sqrt{|qp(p+q)|}\nonumber\\
&&\times\left(\cfie{a}{p+q,t}\fie{a}{q,t}\fie{a}{p,t}+\frac{\fie{a}{-q-p,t}\fie{a}{q,t}\fie{a}{p,t}}{3}+\mbox{h.c.}\right),\label{Model14}
\end{eqnarray}
with the vertex function
\eq{Model15}{
v_{q,p,k}=\kappa_{bc}\sqrt{\tfrac{\pi}{2K}}\left(\frac{qp}{|qp|}+\frac{kp}{|kp|}+\frac{qk}{|qk|}\right)+\kappa_{qp}\sqrt{\tfrac{9K^3\pi}{2}}.}
The interaction \eqref{Model14} describes cubic phonon scattering processes with total momentum conservation. However, not all of the processes contained in $S\sub{Int}$ are resonant, i.e. exactly energy conserving in the sense that $\epsilon_{p+q}=\epsilon_q+\epsilon_p$. As explained by Andreev \cite{andreev80} and pointed out above, the resonant processes lead to a divergence of the self-energy (and the kinetic equation, as we see later) in perturbation theory and are therefore the only relevant terms from a dynamical perspective. The term in Eq.~\eqref{Model14} describing the annihilation (creation) of three phonons can never be resonant. It will therefore play no role in our analysis and we will skip it from now on. For the residual terms, resonance requires $|p+q|=|p|+|q|$. For all momenta $p,q$ fulfilling this condition, the vertex function takes on the value
\eq{Model16}{
v_0\equiv v_{1,1,1}=\sqrt{\frac{9\pi}{2K}}\left(\kappa_{bc}+K^2\kappa_{qp}\right).}
Consequently, instead of taking the full action $S\sub{Int}$, it is sufficient to consider the reduced but resonant phonon interaction
\eq{Model17}{
S\sub{Res}=\frac{v_0}{2\pi}\int_{p,q,t}'\sqrt{|pq(p+q)|}\left(\cfie{a}{p+q,t}\fie{a}{q,t}\fie{a}{p,t}+\mbox{h.c.}\right),
}
where the prime in $\int_{p,q}'$ indicates that the integral runs only over momenta $q,p$ which have the same sign.
Together the quadratic action and the resonant phonon interaction describe the dynamics of the interacting Luttinger model in the phonon basis,
\eq{Model18}{
S=S\sub{TL}+S\sub{Res}.
}
\subsection{Keldysh action}
Non-equilibrium field theory is commonly performed in the Keldysh path integral framework, which is able to deal both with equilibrium and true non-equilibrium situations. To set up the Keldysh path integral, one first doubles the degrees of freedom in the theory by introducing plus and minus fields $\ann{a}{+,q,t}, \ann{a}{-,q,t}$, representing forward and backward time evolution on the Keldysh contour\cite{kamenevbook,AltlandBook}. In this representation, the partition function is determined via
\eq{Model19}{
Z=\int \mathcal{D}[\ann{a}{+},\ann{a}{-},\cre{a}{+},\cre{a}{-}]\ e^{iS_{+}-iS_{-}},}
where $S_{\pm}$ is the phonon action \eqref{Model18} with the replacements $\{\ann{a}{p,t},\cre{a}{p,t}\}\rightarrow\{\ann{a}{\pm,p,t},\cre{a}{\pm,p,t}\}$. The $\pm$-representation contains redundancy, and a technically and physically more appealing representation is found by completing the transformation to the Keldysh representation, introducing classical and quantum fields according to
\eq{Model20}{
\ann{a}{c}=\frac{1}{\sqrt{2}}\left(\ann{a}{+}+\ann{a}{-}\right), \ \ \ \ann{a}{q}=\frac{1}{\sqrt{2}}\left(\ann{a}{+}-\ann{a}{-}\right).}
In the Keldysh representation, the quadratic action is
\eq{Model21}{
S^{(2)}=\frac{1}{2\pi}\int_{t,t',p}\left(\cre{a}{p,t}^c,\cre{a}{p,t}^q\right)\left(\begin{array}{cc} 0& D^R_{p,t,t'}\\D^A_{p,t,t'} & D^{K}_{p,t,t'}  \end{array}\right)\left(\begin{array}{c}\ann{a}{p,t'}^{c}\\ \ann{a}{p,t'}^q\end{array}\right),
}
with the bare inverse retarded/advanced propagators
\begin{eqnarray}
D^R_{p,t,t'}&=&\delta(t-t')\left(i\partial_{t'}-u|p|+i0^{+}\right), \\
D^A_{p,t,t'}&=&\left(D^R_{p,t,t'}\right)^{\dagger}=\delta(t-t')\left(i\partial_{t'}-u|p|-i0^+\right)
\end{eqnarray}
and the Keldysh component of the inverse propagator
\eq{Model24}{
D^K_{p,t,t'}=2i0^+ F(p,t,t').}
Here, $F(p,t,t')$ is the distribution function of the excitations and $0^+$ is the infinitesimal regularization for the quadratic theory \cite{kamenevbook}. In an equilibrium, i.e. time-translational invariant situation, $F(p,t,t')=F(p,t-t')$ and its Fourier transform is the bosonic distribution
\eq{Model25}{
F(p,\omega)=\mbox{coth}\left(\frac{\omega}{2T}\right)=2n_B(\omega)+1}
with the Bose function $n_B(\omega)=\left(e^{\frac{\omega}{T}}-1\right)^{-1}$.
The resonant interactions in the Keldysh representation take on the form
\begin{eqnarray}
S\sub{Res}&=&\frac{v_0}{\sqrt{8}\pi}\int_{p,k,t}' \sqrt{|pk(k+p)|}\ \Big[ 2\cre{a}{k+p,t}^c\ann{a}{k,t}^c\ann{a}{p,t}^q\nonumber\\&&+\cre{a}{k+p,t}^q\left(\ann{a}{k,t}^c\ann{a}{p,t}^c+\ann{a}{k,t}^q\ann{a}{p,t}^q\right)+\mbox{h.c.}\Big].\label{Model26}
\end{eqnarray}

The bare response and correlation functions (retarded, advanced and Keldysh Green's functions) for the phonon degrees of freedom are obtained according to
\begin{eqnarray}
G^R_{q,t,t'}&=&-i \langle \ann{a}{q,t}^c\cre{a}{q,t'}^q\rangle=\left(D^R\right)^{-1}_{q,t,t'}=-i\Theta(t-t')\ e^{-i u|q|(t-t')},\nonumber\\
G^A_{q,t,t'}&=&-i\langle \ann{a}{q,t}^q\cre{a}{q,t'}^c\rangle=\left(D^A\right)^{-1}_{q,t,t'}=i\Theta(t'-t)\ e^{-iu|q|(t-t')},\nonumber\\
G^K_{q,t,t'}&=&-i\langle \ann{a}{q,t}^c\cre{a}{q,t'}^c\rangle=-\left(G^R\circ D^K \circ G^A\right)_{q,t,t'}\nonumber\\
&=&-i \left(2n_B(u|q|)+1\right)\ e^{-iu|q|(t-t')}.\label{Model27}
\end{eqnarray}
Here $...\circ ...$ stands for the convolution in the non-diagonal elements, i.e. the time index, but means multiplication in momentum space.

In the presence of interactions, the Green's functions are modified by the emergence of non-zero self-energies $\Sigma^{R/A/K}$, which replaces the infinitesimal regularization. The corresponding formulas are 
\begin{eqnarray}
G^R_{q,t,t'}&=&\left(D^R-\Sigma^R\right)^{-1}_{q,t,t'},\nonumber\\
G^A_{q,t,t'}&=&\left(D^A-\Sigma^A\right)^{-1}_{q,t,t'},\nonumber\\
G^K_{q,t,t'}&=&-\left(G^R\circ \Sigma^K \circ G^A\right)_{q,t,t'},\label{Model28}
\end{eqnarray}
where the infinitesimal factor $0^+ F$ has been overwritten by the finite Keldysh self-energy $\Sigma^K$. The distribution function $F$ in the presence of interactions is determined by the formula
\eq{Model29}{
G^K_{q,t,t'}=\left(G^R\circ F-F\circ G^A\right)_{q,t,t'}.}
This setting corresponds to the physical situation, in which a system is initialized at time $t=0$ in a Gaussian density matrix $\rho_0$ (Gaussian in the bosonized language). It then evolves in time according to a Hamiltonian $H$, which constitutes the quadratic and cubic terms discussed in Eqs.~\eqref{Model21} and \eqref{Model26}. In the Keldysh setting, this dynamics is expressed in terms of the retarded and advanced quadratic and cubic parts of the action. The initial density matrix enters the action in terms of pure quantum vertices. In the present case, i.e. for purely Gaussian initial conditions, the initial density matrix is completely captured in terms of the distribution function $F$ and Eq.~\eqref{Model24}. However, as has been pointed out recently, the bosonization procedure for interacting fermions or bosons out of equilibrium is in general not that simple, since initial conditions (even if quadratic in the microscopic fermionic or bosonic picture) will, in principle, generate quantum vertices of arbitrary order\cite{Gutman10,Gutman10a,Gutman11,Gutman12,Chernii14}, which have to be taken into account systematically for these cases.
The aim of this work is to determine the self-energies $\Sigma^{R/A/K}$ and the distribution function $F(q,t,t')$ for a system that is driven out of equilibrium and evolves in time, for instance relaxing to an equilibrium state and approaching a bosonic distribution.

This will be done in two parts. First, we show how one determines the self-energies $\Sigma^{R/A/K}$ from a given (non-) equilibrium distribution function $F(q,t,t')$. To this end, we generalize Andreev's self-consistent Born approach to a non-thermal, non-equilibrium situation. Second, we use the kinetic equation approach to determine the time-evolution of the distribution function $F$ in self-consistent Born approximation. Combining these two approaches allows us to determine the time-evolution of both the distribution function of the excitations and the self-energies, which play the role of finite lifetimes of the system's excitations.
\section{Self-energies}\label{sec:Sec3}
The presence of the cubic, resonant interactions $S\sub{Res}$ modifies the phonon response and correlation functions according to Eqs.~\eqref{Model28} by creating finite self-energies $\Sigma^{R/A/K}$. These self-energies are to leading order purely imaginary, leading to a finite decay rate of phonons or, in other words, to a finite phonon lifetime. We will now derive a method to determine these lifetimes for a non-equilibrium problem, where the distribution function $F(q,t,t')$ is time dependent and varies on time scales which are larger than the individual phonon lifetimes. To this end, we first derive the non-equilibrium version of a fluctuation-dissipation relation for the two-point response and correlation functions.
\subsection{Non-equilibrium fluctuation-dissipation relation}
Fluctuation-dissipation relations (FDR) relate the response (i.e. spectral) properties of the system encoded in $G^{R/A}$ to its correlations via the distribution function $F$. A particular example for such a relation is Eq.~\eqref{Model29}. Inverting this equation results in an FDR for the self-energies,
\begin{eqnarray}
\Sigma^K_{q,t,t'}&=&-\left(\left(D^R-\Sigma^R\right)\circ F-F\circ \left(D^A-\Sigma^A\right)\right)_{q,t,t'}\nonumber\\
&=&-i(\partial_t+\partial_{t'})F_{q,t,t'}+\left(\Sigma^R\circ F-F\circ \Sigma^A\right)_{q,t,t'}.\label{SelfEn1}
\end{eqnarray}
For a time-translational invariant system, the first term on the r.h.s. equals zero and due to the identity $\Sigma^A=\left(\Sigma^R\right)^{\dagger}$, Eq.~\eqref{SelfEn1} reduces to the well-known relation in frequency space
\eq{SelfEn2}{
\Sigma^K_{q,\omega}=-2i\ \mbox{Im}\left(\Sigma^R_{q,\omega}\right) \ F_{q,\omega}.}
This is consistent with our initial regularization of the quadratic sector for the case $\Sigma^R=-i0^+$. 

A useful representation for a two-time function $F(q,t,t')$ is to choose Wigner coordinates in time, i.e. defining the forward time $\tau=\frac{t+t'}{2}$ and the relative time $\Delta_t=t-t'$\cite{kamenevbook}. Then one defines $F(q,\Delta_t,\tau)\equiv F(q,\tau+\Delta_t/2,\tau-\Delta_t/2)$ and its Fourier transform
\eq{SelfEn3}{
F(q,\omega,\tau)=\int d\Delta_t \ e^{i\Delta_t\omega} F(q,t,\tau).}
Applying Wigner coordinates and the Wigner transformation \eqref{SelfEn3} to the Keldysh self-energy in Eq.~\eqref{SelfEn2} leads to
\eq{SelfEn4}{
\Sigma^K_{q,\omega,\tau}=-i\partial_{\tau}F_{q,\omega,\tau}+\left(\Sigma^R\circ F-F\circ \Sigma^A\right)_{q,\omega,\tau}.}
Eq.~\eqref{SelfEn4} is an exact expression for the Keldysh self-energy in Wigner representation. A complication arises since the Wigner transform of a convolution is not the product of the corresponding Wigner transforms, in contrast to the ordinary Fourier transform. In fact, one finds
\eq{SelfEn5}{
\left(\Sigma^R\circ F\right)_{q,\omega,\tau}=\Sigma^R_{q,\omega,\tau}\ e^{\frac{i}{2}\left(\overset{\leftarrow}{\partial}_{\tau}\overset{\rightarrow}{\partial}_{\omega}-\overset{\leftarrow}{\partial}_{\omega}\overset{\rightarrow}{\partial}_{\tau}\right)}F_{q,\omega,\tau}.}
Without specific knowledge on the functional behavior of $\Sigma^R$ and $F$, Eq.~\eqref{SelfEn5} is hard to evaluate explicitly. We will now briefly discuss a situation, with two particular approximations, which applies to the present model and for which the analytic evaluation of Eq.~\eqref{SelfEn5} is possible.

\subsubsection{Wigner approximation} For the case of scale separation in the forward and relative time, one can approximate the exponential in \eqref{SelfEn5} by the leading order terms. The product $\partial_{\tau}\partial_{\omega}$ expresses the competition between relative time and forward time dynamics, it is small for slow forward time dynamics and fast relative dynamics\cite{kamenevbook,rammer86}. Comparing the zeroth order term with the first order term in an expansion of the exponential, we find the condition for approximating Eq.~\eqref{SelfEn5} by the zeroth order term to be
\eq{SelfEn6}{
1\gg \left|\frac{\partial_{\omega}\Sigma^R_{q,\omega,\tau}}{\Sigma^R_{q,\omega,\tau}}\right|\left|\frac{\partial_{\tau}F_{q,\omega,\tau}}{F_{q,\omega,\tau}}\right|.}
If this condition is fulfilled, it allows for a separation into fast relative time dynamics and slow forward time dynamics. In general, it is a function of the Luttinger parameters $u, K$, the strength of the nonlinearity $v_0$, as well as of the initial phonon distributions. Therefore, a general criterion for the applicability of the separation of time scales cannot be derived at this point. However, as has been shown in a different context\cite{Heating}, for equilibrium or near to equilibrium initial states, it is fulfilled for temperatures $k\sub{B}T<u\Lambda$ smaller than the Luttinger cutoff. As it turns out, the right hand side of Eq.~\eqref{SelfEn6} is a monotonic function of the quasi-particle distribution $n_q$ and a conservative criterion for the applicability of the Wigner approximation is therefore $n_q<\left(e^{\frac{|q|}{\Lambda}}-1\right)^{-1}$ for all momenta $q$, i.e. the distribution should be smaller than the corresponding distribution at the cutoff temperature.
 We can thus apply the Wigner approximation to the FDR, resulting in
\eq{SelfEn7}{
\Sigma^K_{q,\omega,\tau}=-i\partial_{\tau}F_{q,\omega,\tau}+2i\ \mbox{Im}\left(\Sigma_{q,\omega,\tau}\right)\ F_{q,\omega,\tau}.}
The validity of the FDR in Wigner approximation is a very important result. It is commonly used as the starting point for deriving a kinetic equation for the distribution function in arbitrary dimensions. However, in the present case, we will further simplify the FDR by making use of the fact that we are dealing with resonant interactions in one dimension. 
\subsubsection{Quasi-particle approximation and on-shell FDR}
The major effect of the non-linearities in the action is the emergence of finite phonon lifetimes due to resonant phonon-phonon scattering processes. The resonant character of this interaction -- the fact that for two phonons travelling in the same direction momentum and energy conservation is expressed by the identical $\delta$-constraint -- is the key property of one-dimensional systems with linear dispersion. The resonant contributions dominate the self-energy and the kinetic equation, while the non-resonant processes give only subleading contributions to the lifetimes and the dispersion and have therefore already been eliminated on the basis of the action by using $S\sub{Res}$ instead of $S\sub{Int}$ in Eq.~\eqref{Model18}.

The retarded self-energy is decomposed according to
\eq{SelfEn8}{
\Sigma^R_{q,\omega,\tau}=-i\sigma^R_{q,\tau}+\delta\Sigma^R_{q,\omega,\tau},}
where $\sigma^R_{q,\tau}$ is a {\it positive, frequency independent} function, which varies slowly in forward time $\tau$. For resonant interactions $\delta\Sigma^R_{q,\omega,\tau}=0$ and consequently, the self-energy is frequency independent and purely imaginary. Non-resonant contributions in the interaction lead to $\delta\Sigma^R\neq0$, which however is generally strongly subleading compared to $\sigma^R$.

For the present model, the phonon interactions are RG-irrelevant and only their resonant character allows them to be of non-negligible influence. However, the effect of the interactions on the properties of the phonons will be small and subleading due to the RG-irrelevance. As a result, even in the presence of interactions, the phonons will have a lifetime $\tau^q\sub{ph}=-\mbox{Im}\left(\Sigma^R_q\right)^{-1}$ much larger than their associated coherent time-scale $\frac{1}{u|q|}$, i.e. $\tau\sub{ph}^q\gg \frac{1}{u|q|}$. Consequently, the phonons remain well defined quasi-particles with a spectral function $\mathcal{A}_{q,\omega,\tau}=i\left(G^R-G^A\right)_{q,\omega,\tau}$ sharply peaked at the phonon energy $\omega=u|q|$ \footnote{This underlies the functioning of the linear Luttinger-Liquid theory in equilibrium. In non-equilibrium situations, the RG-irrelevant nature of the interaction leads to a slowly varying density, and therefore the argument holds true even for this case}. For well defined quasi-particles, the self-energies $\Sigma^{R/A/K}$ and the distribution function are evaluated on-shell, i.e. the frequencies are locked according to $\omega=u|q|$, since frequencies $\omega\neq u|q|$ do not contribute to the dynamics.
For resonant interactions, on-shell evaluation is implied and
\eq{SelfEn9}{
\delta\Sigma^{R/A}_{q,\omega,\tau}=0,}
as stated above. This is consistent with the result of Andreev's and later works for equilibrium and holds true for the non-equilibrium case as well. The corresponding decomposition for the Keldysh self-energy (with a convenient prefactor) reads
\eq{SelfEn10}{
\Sigma^K_{q,\omega,\tau}=-2i\sigma^K_{q,\tau}+\delta\Sigma^K_{q,\omega,\tau}\ \ \mbox{ with } \ \delta\Sigma^K_{q,\omega,\tau}=0}
for resonant interactions. Inserting Eqs.~\eqref{SelfEn9}, \eqref{SelfEn10} in the non-equilibrium FDR \eqref{SelfEn7} results in the on-shell, non-equilibrium FDR for resonant interactions
\eq{SelfEn11}{
\sigma^K_{q,\tau}=\partial_{\tau}n_{q,\tau}+\sigma^R_{q,\tau}\left(2n_{q,\tau}+1\right).}
Here we have used the fact that for well defined quasi-particles, the on-shell distribution function $F_{q,\omega=u|q|,\tau}=2n_{q,\tau}+1$ equals the time-dependent phonon density $n_{q,\tau}$ at momentum $q$.

Eq.~\eqref{SelfEn11} is the final form of the non-equilibrium FDR that we will use to set up the kinetic equation in the following section and  to determine the Keldysh self-energy $\sigma^K$ for a system, for which the time-dependent phonon density $n_{q,\tau}$ is known. For the latter case, the only unknown quantity is the retarded self-energy $\sigma^R$ and we can now set up the diagrammatic computation of $\sigma^R$ in the Keldysh non-equilibrium framework.

\subsection{Quasi-particle lifetimes in self-consistent Born approximation}\label{sec:OnShell}
In this section, we perform the self-consistent Born approximation in a diagrammatic representation to obtain the self-energies $\sigma^R_{q,\tau}$. This amounts to an infinite resumation over a certain class of diagrams and cures the divergence of the self-energy occurring in the perturbative diagrammatic approach. In Sec.~\ref{sec:DS}, we demonstrate that the self-energy can be determined exactly in a one-loop computation using Dyson-Schwinger equations and show that the deviation from the self-consistent Born approximation is negligible for many initial states.
\begin{figure}
  \includegraphics[width=8.5cm]{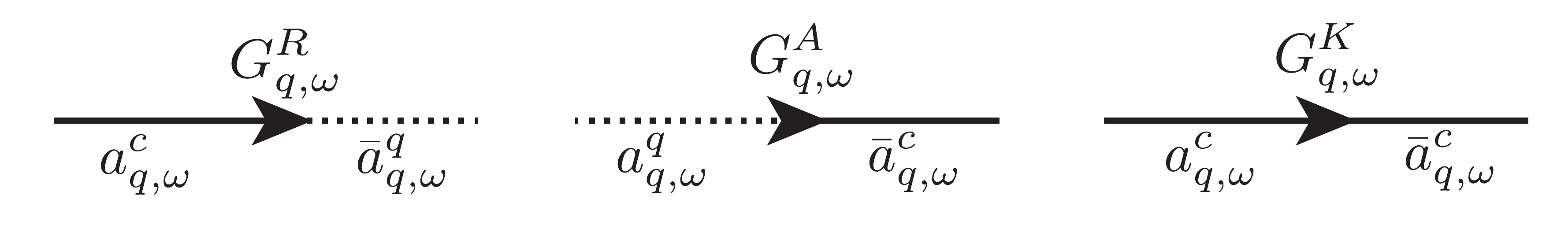}
  \caption{Green's functions in a diagrammatic representation. Full (dotted) lines represent classical (quantum) fields, while ingoing (outgoing) lines represent fields $a$ (hermitian conjugate fields $\bar{a}$). The time index has been omitted}
  \label{fig:Greens}
\end{figure}

\begin{figure}
  \includegraphics[width=8.6cm]{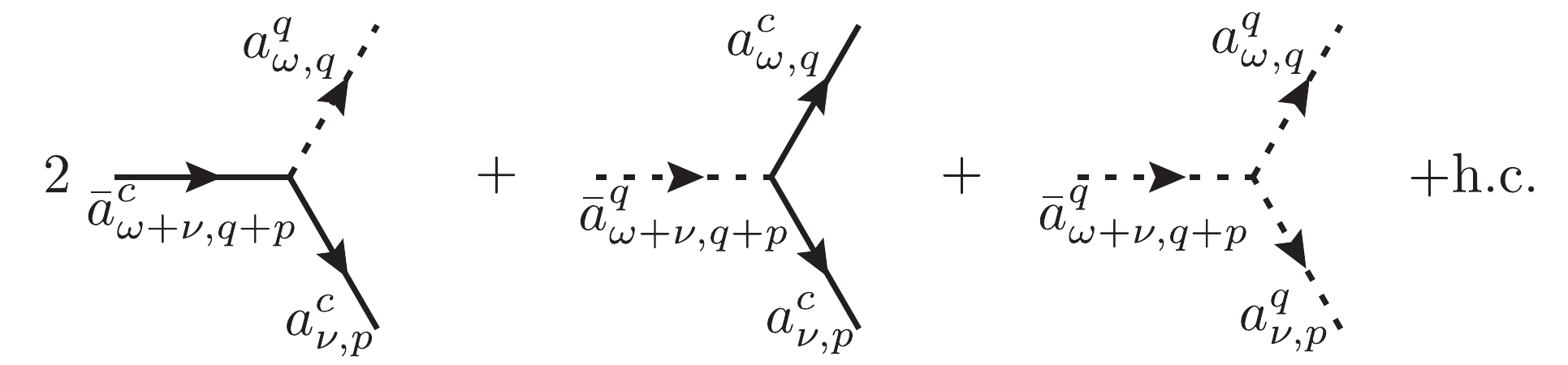}
  \caption{Diagrammatic representation of $S\sub{Res}$, see Eq.~\eqref{Model26}. In total, there exist six different vertices, three as depicted above and three corresponding hermitian conjugates. Each vertex has the prefactor $V(p,q)=\frac{v_0}{\sqrt{8}\pi}\sqrt{pq(p+q)}$.}
  \label{fig:Vertices}
\end{figure}

\begin{figure}
  \includegraphics[width=8.6cm]{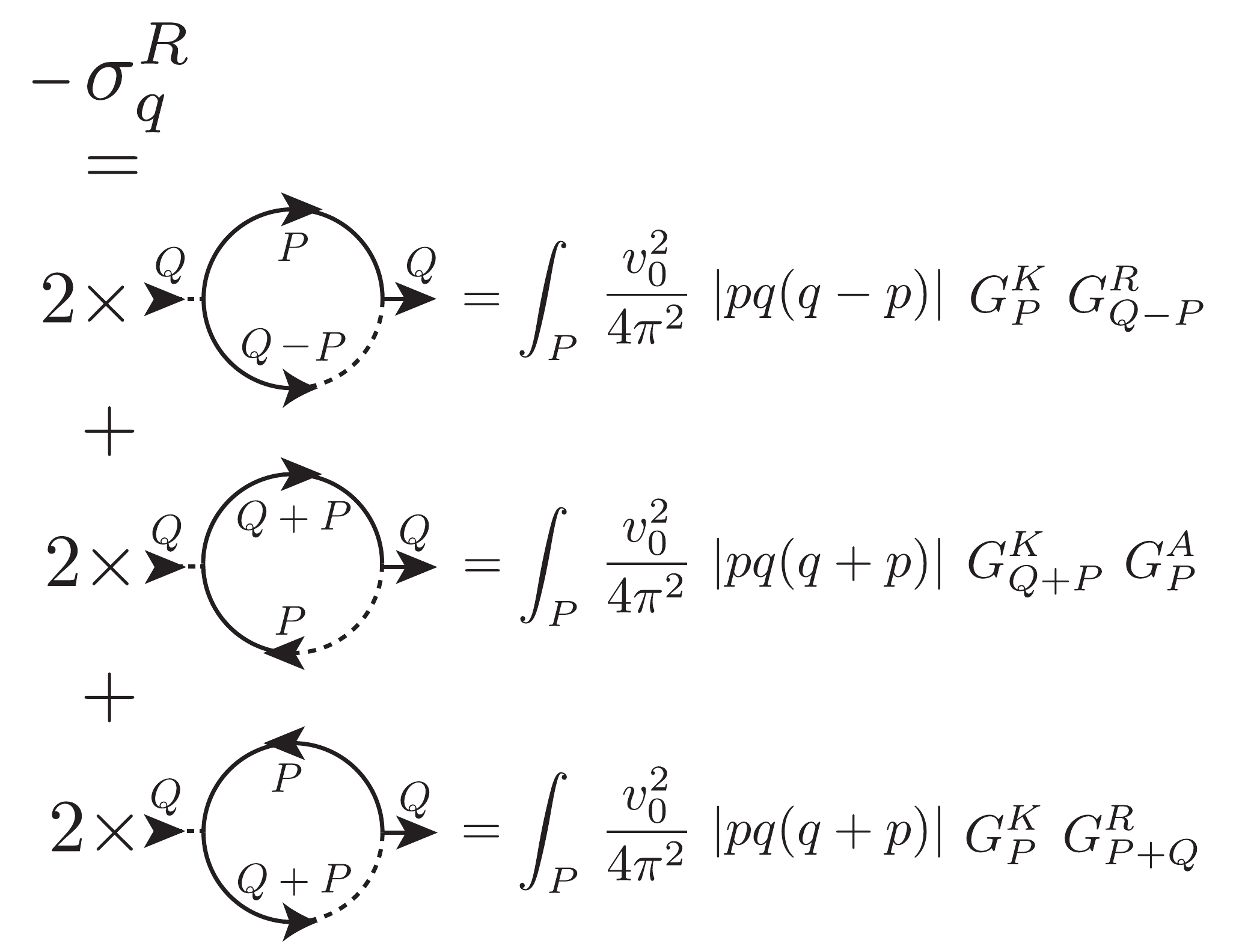}
  \caption{The retarded self-energy is the sum of three distinct diagrams as shown above. The notation is $Q\equiv\{\omega,q\}$, $P=\{\nu,p\}$ with $p,q$ being momenta and $\omega,\nu$ the corresponding frequencies. The momentum integral runs only over momenta that fulfill the resonance condition, which requires $\omega=u|q|$.}
  \label{fig:Diagrams}
\end{figure}

In a diagrammatic approach, classical (quantum) fields are represented by a full (dotted) line, while ingoing lines represent fields $a$ and outgoing lines their complex conjugate $\bar{a}$. This leads to a representation of Green's functions in terms of diagrams as indicated in Fig.~\ref{fig:Greens}, and vertices as depicted in Fig.~\ref{fig:Vertices}.
The retarded self-energy in self-consistent Born approximation can directly be derived by common diagrammatic rules and is depicted in Fig.~\ref{fig:Diagrams}.
Inserting the Keldysh Green's function
\eq{SelfEn12}{
G^K_{q,\omega}=G^R_{q,\omega}\Sigma^K_{q,\omega}G^A_{q,\omega}
}
and the on-shell self-energies $\Sigma^{R/A}_{q,\omega}=\mp i\sigma^R_q, \Sigma^K_{q,\omega}=-2i\sigma^K_q$, we can perform the frequency integration indicated in Fig.~\ref{fig:Diagrams} and find for momenta $q>0$
\begin{eqnarray}
\sigma^R_q&=&v_0^2\left\{\int_{0<p<q}\frac{pq(q-p)\sigma^K_p}{\sigma^R_p\left(\sigma^R_p+\sigma^R_{q-p}\right)}\right.-\int_{q<p}\frac{pq(p-q)\sigma^K_p}{\sigma^R_p\left(\sigma^R_p+\sigma^R_{p-q}\right)}\nonumber\\
&&\left.+\int_{0<p}\frac{pq(p+q)\sigma^K_p}{\sigma^R_p\left(\sigma^R_p+\sigma^R_{p+q}\right)}\right\}.\label{SelfEn13}
\end{eqnarray}
Since the self-energies must be invariant under the transformation $p\rightarrow -p$, one can further simplify Eq.~\eqref{SelfEn13}, ending up with
\eq{SelfEn14}{
\sigma^R_q= v_0^2\int_{0<p}\frac{\sigma^K_p}{\sigma^R_p}\left(\frac{qp(q-p)}{\sigma^R_p+\sigma^R_{q-p}}+\frac{qp(p+q)}{\sigma^R_p+\sigma^R_{p+q}} \right).}
Finally, $\sigma^K$ can be replaced using the FDR \eqref{SelfEn11}, which leads to
\eq{SelfEn15}{
\sigma^R_q= v_0^2\int_{0<p}\left(\frac{\partial_{\tau}n_p}{\sigma^R_p}+2n_p+1\right)\left(\frac{qp(q-p)}{\sigma^R_p+\sigma^R_{q-p}}+\frac{qp(p+q)}{\sigma^R_p+\sigma^R_{p+q}} \right).}
For a given, time dependent distribution function $n_{q,\tau}$, $\sigma^R_{q,\tau}$ is the only unknown function in this equation and has to be determined self-consistently. For a general time-dependent function $n_{q,\tau}$ this has to be done by iterating Eq.~\eqref{SelfEn15} numerically until a self-consistent solution has been found. For the particular case for which $n_{q,\tau}$ shows scaling behavior in a sufficiently large momentum window, one can determine a scaling solution for the self-energy as well and extract the corresponding scaling exponent\cite{andreev80,zwerger06}. We will now briefly discuss the latter case and determine possible scaling solutions for the self-energy, and close the section with a discussion on universal aspects of the scaling solution.

\subsubsection{Scaling solution for the self-energy} For the case when the density $n_{q,\tau}$ is a scaling function, i.e. 
\eq{SelfEn16}{
n_{q,\tau}=a_{\tau} |q|^{\eta_n},
}
it is easy to show that also $\sigma^R_{q,\tau}$ will be a scaling function
\eq{SelfEn17}{
\sigma^R_{q,\tau}=\gamma^R_{\tau} |q|^{\eta^R},}
where the exponent $\eta^R$ and prefactor $\gamma^R_{\tau}$ solely depend on the scaling behavior of $n_{q,\tau}$, i.e. on $\eta_n$ and $a_{\tau}$. In order to show this, we introduce the rescaled self-energy $\tilde{\sigma}^R=\sigma^R/v_0$ and time $\tau=\tilde{\tau}/v_0$, leading to
\eq{SelfEn18}{
\tilde{\sigma}^R_q= \int_{0<p}\left(\frac{\partial_{\tilde{\tau}}n_p}{\tilde{\sigma}^R_p}+2n_p+1\right)\left(\frac{qp(q-p)}{\tilde{\sigma}^R_p+\tilde{\sigma}^R_{q-p}}+\frac{qp(p+q)}{\tilde{\sigma}^R_p+\tilde{\sigma}^R_{p+q}} \right).}
Next, we insert Eqs.~\eqref{SelfEn16}, \eqref{SelfEn17} into \eqref{SelfEn18} and perform the transformation $p\rightarrow q x$, yielding
\begin{eqnarray}
\tilde{\gamma}^R_{\tilde{\tau}}q^{\eta^R}&=&\frac{1}{\tilde{\gamma}^R_{\tilde{\tau}}}q^{4-\eta^R}\int_{0<x}\left[\left(\frac{x(1-x)}{x^{\eta^R}+|1-x|^{\eta^R}}+\frac{x(1+x)}{x^{\eta^R}+|1+x|^{\eta^R}}\right)\right.\nonumber\\
&&\left.\times\left(1+2a_{\tilde{\tau}}(qx)^{\eta_n}+\frac{\partial_{\tilde{\tau}}a_{\tilde{\tau}}}{\tilde{\gamma}^R_{\tilde{\tau}}}(qx)^{\eta_n-\eta^R} \right)\right].\label{SelfEn19}
\end{eqnarray}
The exponent $\eta^R$ is bounded from below and from above according to $1<\eta^R\le 2$. Here, $1<\eta^R$ results from the fact that the interaction is RG irrelevant and the self-energy correction can only be subleading compared to the dispersion $\epsilon_q=u|q|$. The case $\eta^R=2$, i.e. diffusive scaling of the lifetimes is only reached in the zero temperature situation, otherwise one expects superdiffusive behavior due to finite phonon densities with $\eta^R<2$. As a result, Eq.~\eqref{SelfEn19} distinguishes between three regimes:

\underline{1. Low temperature states $(T\approx0)$:} For $n_q\ll1$ and ${\partial_{\tau}n_q\approx0}$ the only term in the second row of Eq.~\eqref{SelfEn19} with a relevant contribution is the constant unity and
\eq{SelfEn20}{
\eta^R=2, \ \ \ \ \tilde{\gamma}^R_{\tau}=\sqrt{I_{2,0}}=\frac{\sqrt{\pi}}{2},}
where the factor $I_{i_1,i_2}$ will be defined in the following.

\underline {2. Finite temperature states:} For $n_q\gg1$ and\\ $q>\mbox{max}\{ \frac{\dot{a}_{\tau}}{2a_{\tau}\tilde{\gamma}_R},\sqrt{\frac{\dot{a}_{\tau}}{2a_{\tau}\tilde{\gamma}_R}}\}$, only the factor proportional to $n_q$ contributes and we have
\eq{SelfEn21}{
\eta^R=2+\frac{\eta_n}{2},\ \ \ \ \tilde{\gamma}^R_{\tau}=\sqrt{2a_{\tau}I_{2+\frac{\eta_n}{2},\eta_n}}.}
For a finite temperature state, $n_q=\frac{T}{u|q|}$ as $|q|\rightarrow0$, such that 
\eq{ThermSelf}{
\tilde{\sigma}^R_q=\sqrt{\frac{2T I_{\frac{3}{2},-1}}{u}}q^{\frac{3}{2}}\approx 0.789 \sqrt{\frac{2\pi T}{u}}q^{\frac{3}{2}}.
}

\underline{3. Non-equilibrium states:} For $n_q\gg1$ and\\ $q<\mbox{min}\{ \frac{\dot{a}_{\tau}}{2a_{\tau}\tilde{\gamma}_R},\sqrt{\frac{\dot{a}_{\tau}}{2a_{\tau}\tilde{\gamma}_R}}\}$, the dominant contribution stems from the time derivative in Eq.~\eqref{SelfEn19} and one finds
\eq{SelfEn22}{
\eta^R=\frac{4+\eta_n}{3},\ \ \ \ \tilde{\gamma}^R=\sqrt[3]{\dot{a}_{\tau}I_{\frac{4+\eta_n}{3},\frac{2\eta_n-4}{3}}}.}
The integral factor is defined as
\eq{SelfEn23}{
I_{i_1,i_2}=\int_{0<x}\left(\frac{x^{1+i_2}(1-x)}{x^{i_1}+|1-x|^{i_1}}+\frac{x^{1+i_2}(1+x)}{x^{i_1}+|1+x|^{i_1}}\right).
}
This gives rise to the possibility of generic non-equilibrium scaling behavior for out-of-equilibrium states, with an exponent $\eta^R$, which is different from the zero and finite temperature cases. A possible scenario of such scaling is discussed in Ref.~\cite{Heating} for a driven system of interacting bosons with a non-equilibrium distribution with $\eta_n=1$ in the infrared.

\begin{figure}
  \includegraphics[width=8cm]{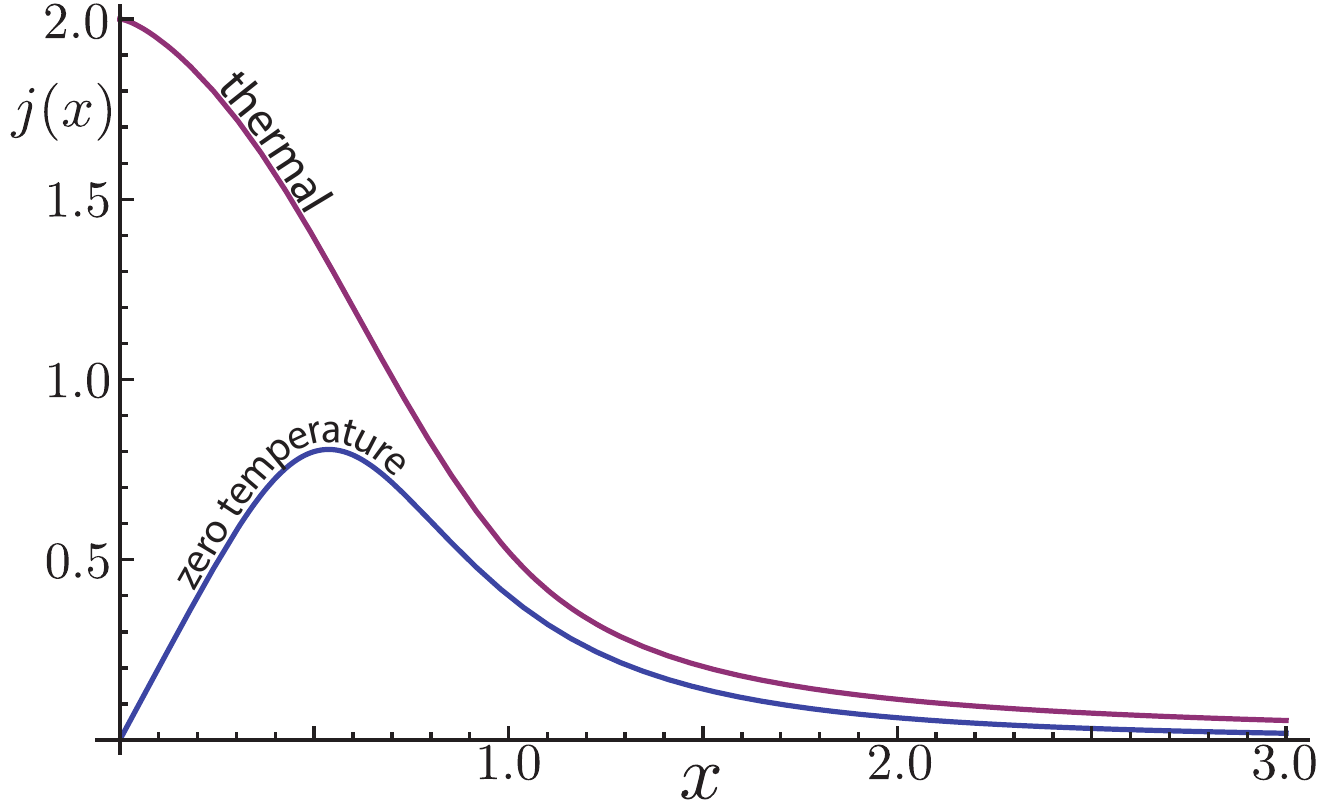}
  \caption{Integrand $j_{i_1,i_2}(x)$ in the self-energy integral $I_{i_1,i_2}=\int_x j_{i_1,i_2}(x)$ as a function of $x=p/q$, where $q$ is the external momentum. The integrand $j(x)$ elucidates the contribution to $\sigma^R_q$ from different momenta and shows clearly that the self-energy at momentum $q$ is dominated by the behavior of $\sigma^R_p$ for $p<q$. For physically relevant distribution functions, with the property $n_q\rightarrow0$ for $q\rightarrow\Lambda$, the exponent $\eta^R\rightarrow2$ for $x\rightarrow\infty$, which leads to a decay of the integrand $\sim x^{-3}$ and a well-defined self-energy integral.}
  \label{fig:Integrand}
\end{figure}
The results discussed above include the case of finite and zero temperature equilibrium. For the latter $n_q=0$ and therefore
$\eta^R=2, \gamma^R_{\tau}=v_0\sqrt{\frac{\pi}{4}}$, while for finite temperature $n_q\approx \frac{T}{u|q|}$ and $\partial_{\tau}n_{q}=0$ and consequently $\eta^R=3/2, \gamma^R_{\tau}=0.789v_0\sqrt{\frac{2 \pi T}{u}}$. These are the mentioned results obtained for the zero and finite temperature equilibrium cases\cite{andreev80,zwerger06,affleck06}.
\subsubsection{Insensitivity of the self-energy to UV-behavior}
The insensitivity of the above results to the  behavior of the model in the ultraviolet (UV) regime and therefore the non-universal properties of the system is guaranteed by the structure of the vertex and the diagrams in Fig.~\ref{fig:Diagrams}. The self-energies $\sigma^R_q$ are dominated by loop momenta $p<q$ below the external momentum $q$. Therefore the non-universal behavior in the UV does not enter the self-energies. This is emphasized in Fig.~\ref{fig:Integrand}, where the integrand
\eq{SelfEn24}{
j_{i_1,i_2}(x)=\frac{(1-x)x^{1+i_2}}{|1-x|^{i_1}+x^{i_1}}+\frac{(1+x)x^{1+i_2}}{(1+x)^{i_1}+x^{i_1}}}
in $I_{i_1,i_2}=\int_x j_{i_1,i_2}(x)$, $x=p/q$, is plotted for the case of thermal and zero temperature scaling. It is evident that for ingoing momenta $q$ only momenta $p<q$ contribute, which show the same scaling behavior in the distribution function. Therefore the scaling solutions for the self-energy are robust against modifications of the density and the model itself when approaching the UV.
\begin{figure}
  \includegraphics[width=8.6cm]{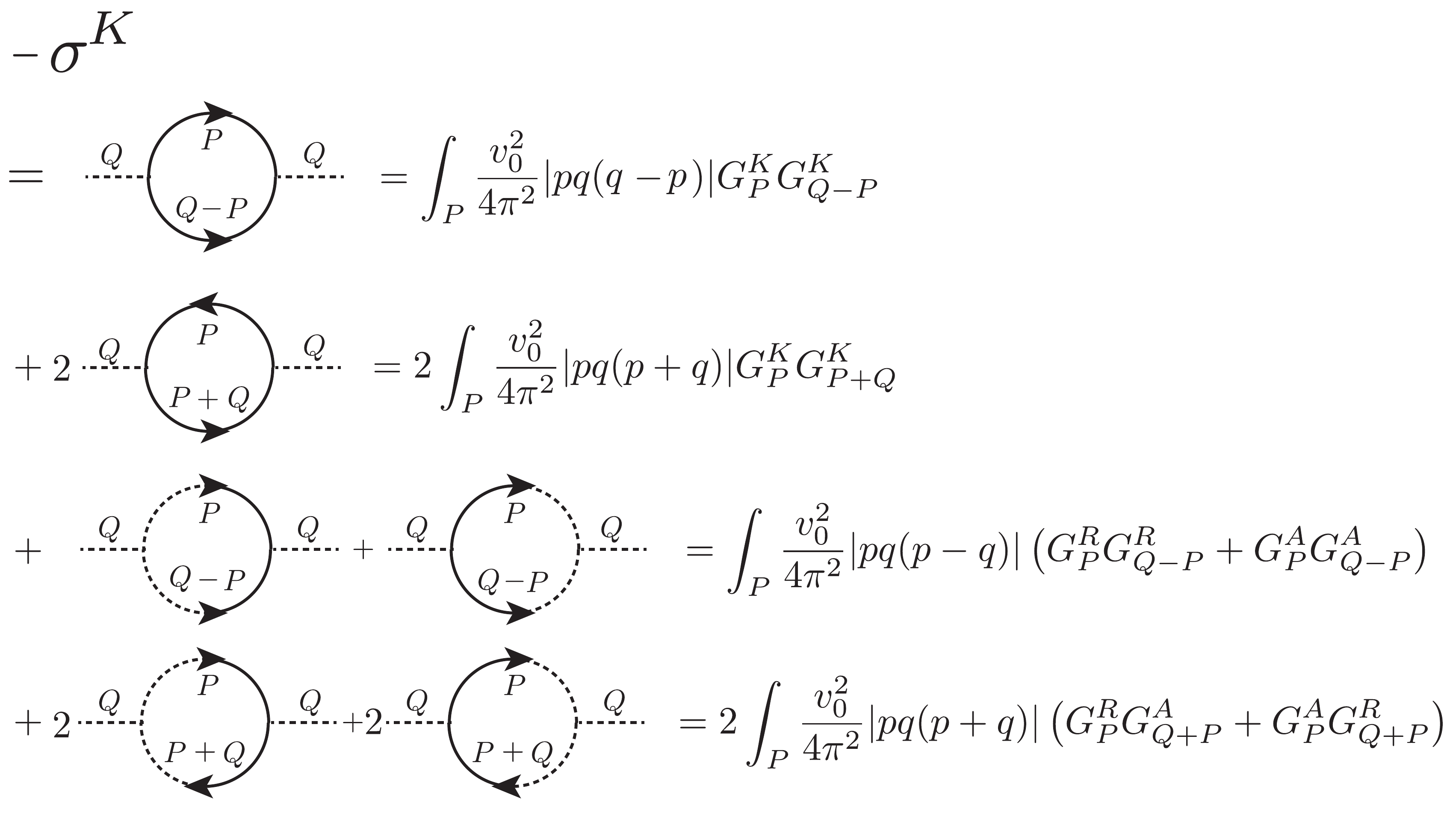}
  \caption{Diagrammatic representation of the Keldysh self-energy $\sigma^K_q$ in self-consistent Born approximation. The index $Q=\{u|q|\pm i\sigma^R_q,q\}$ represents the on-shell frequency and momentum relation ($+$ for retarded, $-$ for advanced Green's functions), while $P=\{\nu,p\}$ is the inner loop frequency and momentum and $\int_{P}=\int_{\nu,p}$. Integration over resonant processes only is implied.
}
  \label{fig:DiagramsK}
\end{figure}

\section{Kinetic Equation for the Phonon Density}\label{sec:Sec4}
In this section, we will derive a kinetic equation for the phonon density in the case of resonant interactions. A kinetic equation describes the time-evolution of the distribution function generated by the Keldysh and retarded/advanced self-energies \cite{kamenevbook,mitrarosch,Tavora13}, which can then often be evaluated in the perturbative Born approximation. In the present case, due to the fact that the interactions are resonant, the kinetic equation perturbation theory diverges and we have to use a self-consistent Born approach. Therefore, we evaluate the self-energy diagrams with full Green's functions, as in the previous section, which leads to an effective vertex correction for the kinetic equation and a time-evolution linear in the interaction parameter $v_0$.

The time-evolution of $n_{q,\tau}$ is determined by the solution of the on-shell FDR in Eq.~\eqref{SelfEn11}, which after rearrangement reads
\eq{Kinetic1}{
\partial_{\tau}n_q=\sigma^K_{q,\tau}-\left(2n_{q,\tau}+1\right)\sigma^R_{q,\tau}.}
Again, the retarded self-energy is determined via a diagrammatic approach, where the corresponding diagrams are shown in Fig.~\ref{fig:Diagrams}. However, in contrast to the previous section, we also use a diagrammatic approach to determine the Keldysh self-energy $\sigma^K$. As a consequence, we derive a non-linear differential equation for the distribution function.
The diagrammatic representation of the Keldysh self-energy is depicted in Fig.~\ref{fig:DiagramsK}.

For the full Green's functions, we use the results from the previous section

\eq{Kinetic2}{
G^{R/A}_{q,\omega,\tau}=2\pi\left(\omega-u|q|\pm i\sigma^R_{q,\tau}\right)^{-1}}
and the relation
\eq{Kinetic3}{
G^K_{q,\omega,\tau}=G^R_{q,\omega,\tau}F_{q,\omega,\tau}-F_{q,\omega,\tau}G^A_{q,\omega,\tau}=\frac{-8\pi^2i\sigma^R_{q,\tau}\left(2n_{q,\tau}+1\right)}{\left(\omega-u|q|\right)^2+\left(\sigma^R_{q,\tau}\right)^2}.}
In Eq.~\eqref{Kinetic3}, the first equality holds in Wigner approximation, while the second equality results from the quasi-particle approximation, both discussed in the previous section.

The frequency integration in the diagrammatic representation can be performed analytically and yields the kinetic equation (omitting time index)
\begin{eqnarray}
\partial_{\tau}n_{q}\hspace{-0.1cm}&=&\hspace{-0.05cm}2v_0^2\hspace{-0.1cm}\int_{0<p<q}\tfrac{pq(q-p)}{\sigma^R_q+\sigma^R_p+\sigma^R_{q-p}}\left(n_pn_{q-p}-n_q\left(1+n_p+n_{q-p}\right)\right)\nonumber\\
&&+4v_0^2\hspace{-0.1cm}\int_{0<p}\tfrac{pq(q+p)}{\sigma^R_q+\sigma^R_p+\sigma^R_{q+p}}\left(n_{p+q}\left(n_q+n_p+1\right)-n_qn_p    \right).\phantom{dd} \label{Kinetic4}
\end{eqnarray}
After transforming to dimensionless variables $\sigma^R=v_0\tilde{\sigma}^R$, $\tau=\frac{\tilde{\tau}}{v_0}$, we finally arrive at 
\begin{eqnarray}
\partial_{\tilde{\tau}}n_q\hspace{-1mm}&=&\hspace{-1.5mm}\int_{0<p<q}\tfrac{2pq(q-p)}{\tilde{\sigma}^R_q+\tilde{\sigma}^R_p+\tilde{\sigma}^R_{q-p}}\left(n_pn_{q-p}-n_q\left(1+n_p+n_{q-p}\right)\right)\ \ \nonumber\\
&+&\hspace{-1.5mm}\int_{0<p}\hspace{-0.8mm}\tfrac{4pq(q+p)}{\tilde{\sigma}^R_q+\tilde{\sigma}^R_p+\tilde{\sigma}^R_{q+p}}\left(n_{p+q}\left(n_q+n_p+1\right)-n_qn_p    \right).\label{Kinetic5}
\end{eqnarray}
This is the kinetic equation for the phonon density in self-consistent Born approximation.

Comparing this equation to Eq.~\eqref{SelfEn18}, one finds that the rescaled self-energy $\tilde{\sigma}^R_q$ and therefore also the kinetic equation for a rescaled time $\tilde{\tau}=v_0\tau$ only depend on the phonon distribution $n_{q,\tilde{\tau}}$ and is independent of all possible microscopic details that may enter $u, K, v_0$ in the model. As a result, the dynamics in the rescaled variables is very generic for an interacting Luttinger Liquid and only depends on the initial distribution function $n_{q,\tau=0}$ with which the system is initialized.

The typical time-scale for the kinetic equation in the original variables is $\tau\sub{typ}=\frac{1}{v_0}$, i.e. linear in the vertex $v_0$. This is a non-perturbative effect resulting from the resonant interactions. Since two vertices enter the one-loop diagrams, one might naively (or in perturbation theory) expect that the typical time scale is given by the square of the non-linearity. However, this is normalized by the self-energies, which are proportional to $v_0$ and required to regularize the kinetic equation.

\begin{figure}[h]
  \includegraphics[width=8.6cm]{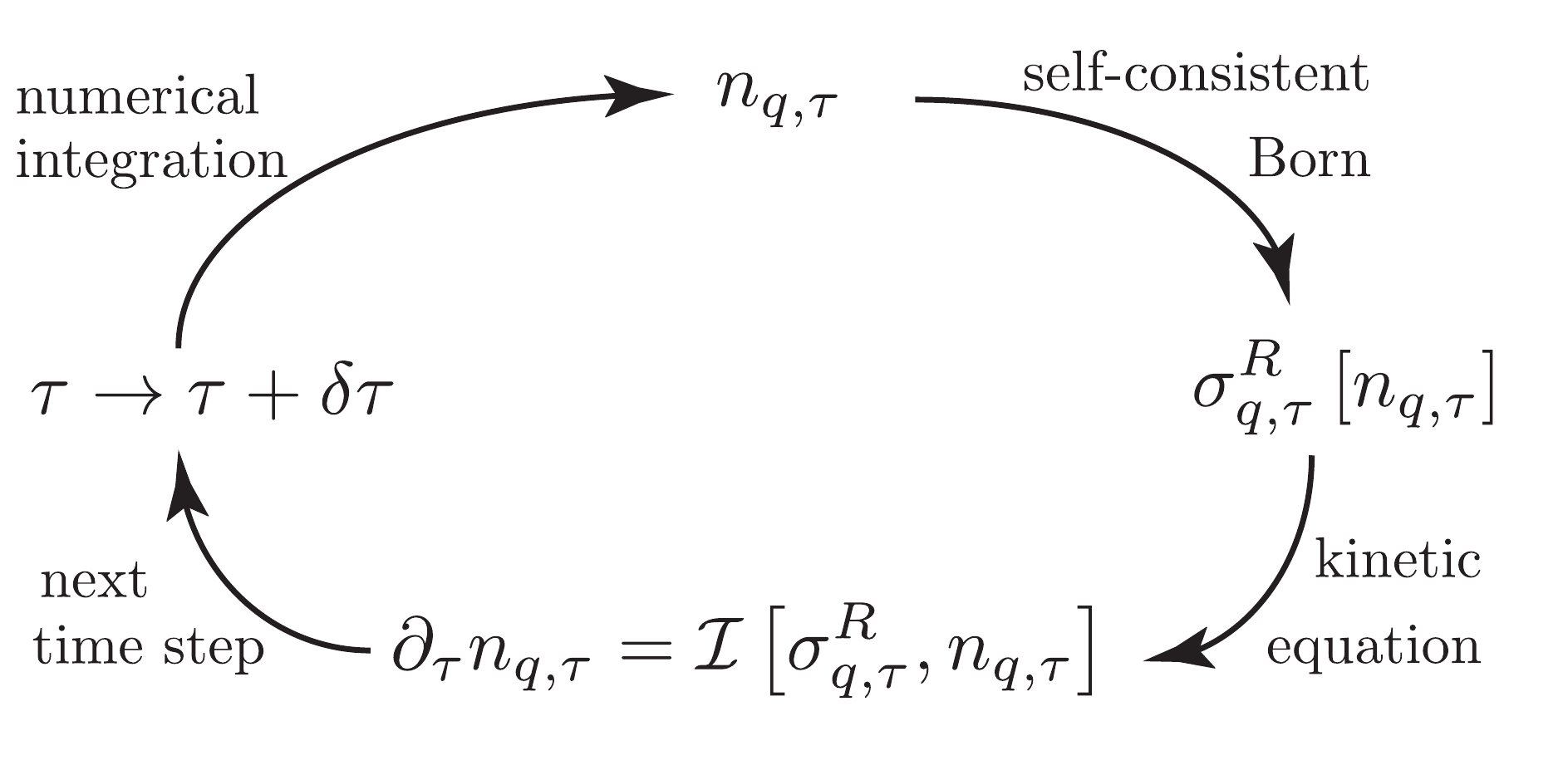}
  \caption{Schematic illustration of the iteration process to determine the time-dependent phonon density $n_{q,\tau}$. For a given time $\tau$, the self-energy $\sigma^R_{q,\tau}$ is determined via the self-consistent Born approximation according to Eq.~\eqref{SelfEn18}. Subsequently, the time-derivative of $n_{q,\tau}$ is computed via the kinetic equation \eqref{Kinetic5}. Using a Runge-Kutta solver for numerical differential equations, the density $n_{q,\tau+\delta\tau}$ can be computed and used as the starting point for the next iteration.}
  \label{fig:Iteration}
\end{figure}

In the kinetic equation in \eqref{Kinetic5}, still the self-energies $\tilde{\sigma}^R$ occur. In principle, one could again replace them by a diagrammatic expression, which would give rise to an infinite hierarchy. However instead of doing so, we use an iterative process in which, for a given time $\tau$, we determine self-consistently the self-energy $\sigma^R_{q,\tau}$. This result is then used to determine the r.h.s. of the kinetic equation, to subsequently compute the distribution function in the next time step $\tau+\delta\tau$. This procedure is illustrated in Fig.~\ref{fig:Iteration} and allows us to compute the non-equilibrium dynamics of an interacting Luttinger Liquid, which may be initialized in a non-thermal state.

We will now close this section by discussing two limiting cases where analytic results become available. First, the kinetic equation for small external momenta $q$, and second the kinetic equation for a phonon distribution $n_{q,\tau}=n_{q,T}+\delta n_{q,\tau}$ close to an equilibrium distribution $n_{q,T}$.

\subsection{Kinetic equation for small external momenta}
For small external momenta, the kinetic equation \eqref{Kinetic5} can be brought into an even simpler form, explicitly revealing the evolution of $n_{q,\tau}$ for small $q$. In this case, the first integral in \eqref{Kinetic5} covers only a very small momentum region and is proportional to $q^{4-\eta^R}$, $1<\eta^R\le 2$. As a result, it is negligible compared to the second integral. On the other hand, in the second integral summations including $q$ can be replaced according to $p+q\approx p$, $\sigma^R_q+\sigma^R_p\approx \sigma^R_p$. The kinetic equation then simplifies to
\eq{Kinetic6}{
\partial_{\tilde{\tau}}n_{q,\tilde{\tau}}\overset{q\ll1}{=}|q|\int_{0<p}\frac{2p^2}{\tilde{\sigma}^R_{p,\tilde{\tau}}}n_{p,\tilde{\tau}}\left(1+n_{p,\tilde{\tau}}\right)=|q|\mathcal{I}_{\tilde{\tau}},
}
where $\mathcal{I}_{\tau}$ is a time dependent but momentum independent functional. The phonon density becomes
\eq{Kinetic7}{
n_{q,\tilde{\tau}}\overset{q\ll 1}{=}n_{q,\tilde{\tau}=0}+|q| \int_{0<t<\tilde{\tau}}\mathcal{I}_t}
for sufficiently small momenta. Crucially, the change is linear in momentum $q$ and for $q=0$ the density is time independent, and for all times pinned to its initial value. This is an exact result, which can be traced back to particle number conservation in the underlying microscopic model \cite{mahan}.
\subsection{Relaxation close to equilibrium}
A stationary solution of the kinetic equation $\partial_{\tau}n_{q,\tau}=0$ is given by the Bose distribution function 
\eq{Kinetic8}{
n_{q,\tau}=n\sub{B}(u|q|,T)=\left(e^{u|q|/T}-1\right)^{-1}}
for arbitrary temperature $T$. Sometimes one is interested in the relaxation of the distribution function close to equilibrium
\eq{Kinetic9}{
n_{q,\tau}=n\sub{B}(u|q|,T)+\delta n_{q,\tau},}
where the variation $\delta n\ll n\sub{B}$. For small momenta $u|q|\ll T$, $n\sub{B}(u|q|,T)=\frac{T}{u|q|}$ and we can expand the kinetic equation in the variation $\delta n$. The zeroth order part solves the kinetic equation, and the leading order contribution stems from the first order of the expansion. 
After eliminating negligible terms, it reads
\begin{eqnarray}
\partial_{\tilde{\tau}}\delta n_q=-\frac{2Tq^2\delta n_q}{u}&&\ \left(\int_{0<p<q}
\frac{1}{\tilde{\sigma}^R_q+\tilde{\sigma}^R_p+\tilde{\sigma}^R_{q-p}}\right. \nonumber\\
&&+\left.\int_{0<p}\frac{2}{\tilde{\sigma}^R_q+\tilde{\sigma}^R_p+\tilde{\sigma}^R_{q+p}}
\right).\label{Kinetic10}\end{eqnarray}

For a distribution close to thermal equilibrium, the self-energy will take on the thermal form \eqref{ThermSelf} and
\eq{Kinetic11}{
\partial_{\tilde{\tau}}\delta n_q\approx-\alpha\sub{T}\ \delta n_q \sqrt{\frac{2\pi T}{u}}q^{\frac{3}{2}}=-\alpha\sub{T}\tilde{\sigma}^R_q\delta n_q,}
where $\tilde{\sigma}^R_q$ is the thermal self-energy and $\alpha\sub{T}\approx1.1$ is a universal number. This result can also be seen as an expansion of Eq.~\eqref{Kinetic1} in $\delta n_q$, which is
\eq{Kinetic12}{
\partial_{\tilde{\tau}}\delta n_q=-2\delta n_q\tilde{\sigma}^R_q+\frac{\partial}{\partial\delta n_q}\left.\left(\tilde{\sigma}^K_q-\left(2n\sub{B}+1\right)\tilde{\sigma}^R_q\right)\right|_{\delta n_q=0}\delta n_q.
}
The second term thus gives a correction to the simple factor of $2$ in Eq.~\eqref{Kinetic11}. We thus obtain a nonperturbative estimate for the relaxation time of the interacting Luttinger Liquid
\eq{RelaxTime}{
\tau_q=\frac{0.868}{v_0}\sqrt{\frac{u}{2\pi T}}q^{-\frac{3}{2}}
}
reflecting the very slow asymptotic approach to equilibrium of the long-wavelength modes. It is very similar to the lifetime of a single thermal phonon (cf. Eq.~\eqref{ThermSelf}), only modified by the prefactor $\alpha\sub{T}=1.1$. This modification arises due to in-scattering processes of excitations $p\neq q$ scattering into the mode $q$. Since the main relaxation process is caused by out-scattering processes, the above result is quite intuitive and supports the statement that relative time dynamics $\sim \frac{1}{\epsilon_q}$ are fast compared to forward time dynamics $\sim\tau_q$.

\section{Kinetic Equation and Diagrams in presence of anomalous densities}\label{sec:Sec5}
In this section, we consider the effect of non-zero anomalous (off-diagonal) phonon density, i.e. $\cre{a}{q}\cre{a}{-q}\neq0$ on the diagonal kinetic equation and self-energy, and derive the kinetic equation for the anomalous density. Off-diagonal phonon densities can be populated due to external perturbations, as for instance density modulation due to a Bragg beam or a global interaction quench \cite{cazalillareview,cazalilla06}, and their impact on the kinetics and non-equilibrium dynamics is therefore non-negligible.
Summarizing the results of this section, due to the structure of the resonant interactions, the kinetic equation for the diagonal phonon density and the diagonal retarded/advanced self-energies are not modified in the presence of anomalous densities and remain unchanged compared to Eqs.~\eqref{Kinetic5} and \eqref{SelfEn15}. In contrast, the kinetic equation for the anomalous densities is fed by the normal occupations, cf. Eq.~\eqref{OffDiag19}.

For a generic equilibrium situation and for certain realizations of systems brought out of equilibrium, the anomalous (off-diagonal) phonon density and consequently the corresponding response and correlation functions are zero, i.e.
\eq{OffDiag1}{
\langle \ann{a}{q,t}^{\alpha}\ann{a}{-q,t'}^{\alpha'}\rangle=0,}
where $\alpha,\alpha'=c,q$ represent classical or quantum indices. However, when a system is driven out of equilibrium, it is possible to populate off-diagonal terms. A simple situation for which this is indeed the case is an interaction quench in a one dimensional quantum fluid, where the off-diagonal densities are non-zero after the quench and lead to non-equilibrium dynamics even in the absence of phonon-phonon scattering processes \cite{cazalilla06}.

In order to deal with the situation of anomalous densities, we first have to modify the FDR accordingly. The Keldysh Green's function in the presence of off-diagonal terms, expressed in Nambu space, is
\eq{OffDiag2}{
G^K_{q,t,t'}=\left(\begin{array}{cc}g^K_{q,t,t'} & h^K_{q,t,t'}\\
-\left(h^K_{q,t,t'}\right)^*& g^K_{-q,t',t}\end{array}\right)=-i \left(\begin{array}{cc}\langle \ann{a}{q,t}^c \cre{a}{q,t'}^c\rangle 
& \langle \ann{a}{q,t}^c\ann{a}{-q,t'}^c\rangle\\
\langle\cre{a}{-q,t}^c\cre{a}{q,t'}^c\rangle &\langle \cre{a}{-q,t}^c\ann{a}{-q,t'}^c\rangle
\end{array}\right).}
For the quadratic theory in the absence of phonon scattering, i.e. $S=S\sub{TL}$ only, the Keldysh Green's function can be evaluated explicitly. In a operator representation, it reads 
\eq{anticom}{
G^K_{q,t,t'}=-i \left(\begin{array}{cc}\langle\{ a^{\phantom{\dagger}}_{q,t}, a^{\dagger}_{q,t'}\}\rangle 
& \langle\{ a^{\phantom{\dagger}}_{q,t},a^{\phantom{\dagger}}_{-q,t'}\}\rangle\\
\langle \{ a^{\dagger}_{-q,t},a^{\dagger}_{q,t'}\}\rangle &\langle \{a^{\dagger}_{-q,t},a^{\phantom{\dagger}}_{-q,t'}\}\rangle
\end{array}\right),}
with the anti-commutator $\{\cdot , \cdot\}$. In Wigner representation it is
\eq{OffDiag3}{
G^K_{q,\omega,\tau}=-i\left(\begin{array}{cc} \delta(\omega-\epsilon_q)(2n_q+1) & \delta(\omega)2m_q e^{-i2\epsilon_q\tau}\\ \delta(\omega) 2m^*_q e^{i2\epsilon_q\tau} & \delta(\omega+\epsilon_q) (2n_{-q}+1)
\end{array}\right),}
where $m_q$ is the anomalous phonon density ($m_q=|\langle a_q a_{-q}\rangle|$ in an operator representation in terms of annihilation operators $a$). In the quadratic theory, both the normal and the anomalous densities are constants of motion. 

The Keldysh Green's function in \eqref{OffDiag3} has two essential drawbacks. For the case of non-zero but slowly (compared to $\epsilon_q$) varying anomalous density $m_q$, the off-diagonal terms of $G^K$ are not slow but oscillate with the fastest scale in the problem, i.e. $t\sub{typ}=\frac{1}{2\epsilon_q}$. The Wigner approximation is therefore not applicable for the off-diagonal terms of $G^K$. Furthermore, the Keldysh Green's function in frequency representation is peaked at three different frequencies, $\omega=(\epsilon_q, 0,-\epsilon_q)$. Both prevents an FDR in the form of Eq.~\eqref{SelfEn7} to exist in this representation.

In order to cure this problem, we switch to a rotating frame by introducing the fields $\ann{\alpha}{q,t}=\ann{a}{q,t}e^{i\epsilon_qt}, \cre{\alpha}{q,t}=\cre{a}{q,t}e^{-i\epsilon_qt}$, which modifies the quadratic action according to
\eq{OffDiag4}{
S^{(2)}=\int_{t,p}\left(\cre{\alpha}{p,t}^c,\cre{\alpha}{p,t}^q\right)\left(\begin{array}{cc} 0& i\partial_t+i0^+\\ i\partial_t-i0^+ & 2i0^+\coth\left(\frac{\omega}{2T}\right)  \end{array}\right)\left(\begin{array}{c}\ann{\alpha}{p,t}^{c}\\ \ann{\alpha}{p,t}^q\end{array}\right).}
The resonant phonon interaction is invariant under the transformation
\begin{eqnarray}
S\sub{Res}&=&\frac{v_0}{\sqrt{2}}\int_{p,k,t}' \sqrt{|pk(k+p)|}\ \Big[ 2\cre{\alpha}{k+p,t}^c\ann{\alpha}{k,t}^c\ann{\alpha}{p,t}^q\nonumber\\&&+\cre{\alpha}{k+p,t}^q\left(\ann{\alpha}{k,t}^c\ann{\alpha}{p,t}^c+\ann{\alpha}{k,t}^q\ann{\alpha}{p,t}^q\right)+\mbox{h.c.}\Big].\label{OffDiag5}
\end{eqnarray}
since the phase $e^{it\left(\epsilon_{p+k}-\epsilon_k-\epsilon_p\right)}=1$ in the case of resonance, i.e. for $|k+p|=|k|+|p|$.
The corresponding correlation function in Nambu space and Wigner coordinates after the rotation is 
\eq{OffDiag6}{
\tilde{G}^K_{q,\omega,\tau}=-i\delta(\omega)\left(\begin{array}{cc} 2n_q+1 & 2m_q\\ 2m_q^*& 2n_{-q}+1\end{array}\right),}
while the bare retarded Green's function reads
\eq{OffDiag7}{
\tilde{G}^R_{q,\omega,\tau}=\left(\begin{array}{cc}\frac{1}{\omega+i0^+} & 0\\ 0 & \frac{1}{-\omega-i0^+} \end{array}\right)=\sigma_z \frac{1}{\omega+i0^+},
}
where $\sigma_z$ is the Pauli matrix. Respecting the symplectic structure in bosonic Nambu space, the FDR in the presence of off-diagonal densities is
\eq{OffDiag8}{
\tilde{G}^K_{q,\omega,\tau}=\left(\tilde{G}^R\circ\sigma_z\circ \tilde{F}-\tilde{F}\circ \sigma_z \circ\tilde{G}^A\right)_{q,\omega,\tau}=-i\delta(\omega)\tilde{F}_{q,\omega,\tau}.}
Here $\tilde{F}$ is the physical distribution function for the phonons, with the on-shell value
\eq{OffDiag9}{
\tilde{F}_{q,\omega=0,\tau}=\left(\begin{array}{cc} 2n_{q,\tau}+1 & 2m_{q,\tau}\\ 2m_{q,\tau}^*& 2n_{-q,\tau}+1\end{array}\right).}
Both the transformation to a rotating frame by switching from $\{\ann{a}{q}, \cre{a}{q}\}$ to $\{\ann{\alpha}{q},\cre{\alpha}{q}\}$ as well as the symplectic factors $\sigma_z$ in Eq.~\eqref{OffDiag8} are necessary modifications in order to obtain an FDR with a physically relevant distribution function $\tilde{F}$. The latter should consist of diagonal and anomalous densities that are slowly varying in time and reproduce the matrix structure of $\tilde{G}^K$.

Inversion of Eq.~\eqref{OffDiag8} yields the kinetic equation 
\eq{OffDiag9}{
i\partial_{\tau}F_{q,\omega,\tau}=\sigma_z\Sigma^R_{q,\omega,\tau}F_{q,\omega,\tau}-F_{q,\omega,\tau}\Sigma^A_{q,\omega,\tau}\sigma_z-\sigma_z\Sigma^K_{q,\omega,\tau}\sigma_z.}

In the absence of off-diagonal terms in both self-energies and the distribution function, this equation reduces to the ordinary kinetic equation discussed in the previous section.
We will now set up the diagrammatic computation of the self-energies $\Sigma^{R/A/K}_{q,\omega,\tau}$ in order to derive the kinetic equation in the presence of anomalous phonon densities. 
\subsection{Diagrammatics for off-diagonal terms}
In the presence of off-diagonal densities, one can no longer generally exclude non-zero off-diagonal self-energies and consequently non-zero off-diagonal Green's functions from the action. In this section we set up the diagrammatic computation of the self-energies in Nambu space. We obtain two key results, which crucially rely on the resonant character of the interaction. {\it First}, in the retarded/advanced sector, the off-diagonal self-energies are exactly zero even in the presence of anomalous densities. {\it Second}, we compute the off-diagonal self-energies in the Keldysh sector, which are different from zero.

\begin{figure}
  \includegraphics[width=8.6cm]{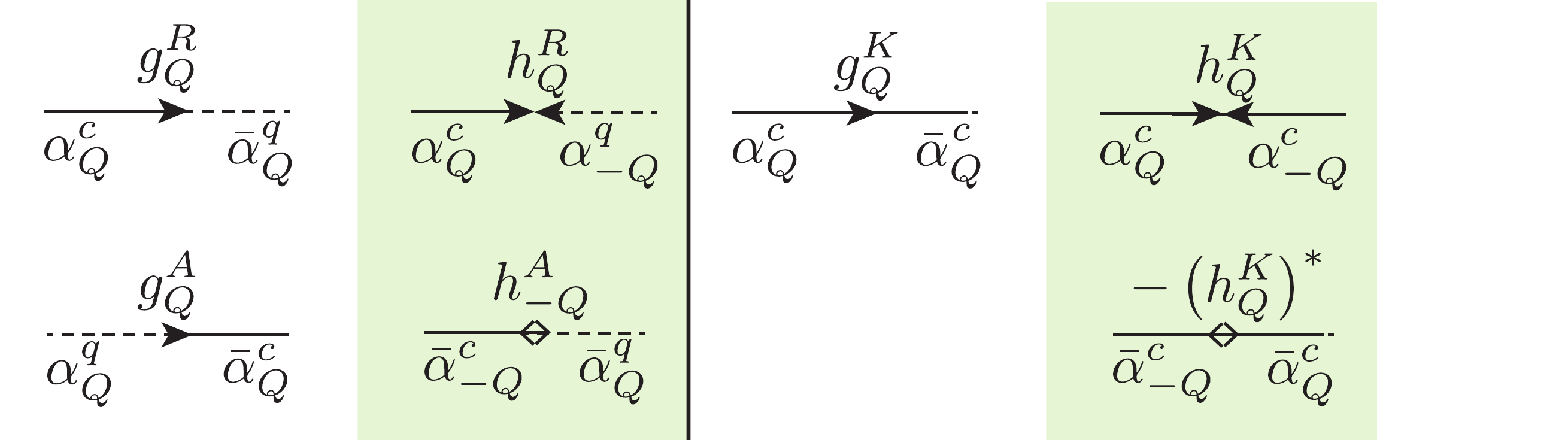}
  \caption{Diagrammatic representation of the diagonal and off-diagonal retarded/advanced and Keldysh Green's functions. In the presence of anomalous densities, the off-diagonal Keldysh components will be generally non-zero, $h^K\neq 0$. The retarded and advanced Green's functions only differ from zero if a finite off-diagonal self-energy is present, cf. Eq.~\eqref{OffDiag11}.}
  \label{fig:OffGreens}
\end{figure}
The retarded Green's function in Nambu space is (we use a general index $Q=(q,\omega,\tau)$, $-Q=(-q,-\omega,\tau)$ for this paragraph)
\begin{eqnarray}
G^R_{Q}&=&\left(\begin{array}{cc}g^R_Q & h^R_Q\\ h^A_{-Q} & g^A_{-Q}
\end{array}\right)=\left(\begin{array}{cc}\omega-\Sigma^R_Q & -\Gamma^R_Q\\ -\Gamma^A_{-Q} & -\omega-\Sigma^A_{-Q}
\end{array}\right)^{-1}\label{OffDiag11}\\
&=&\frac{1}{(\omega-\Sigma^R_Q)(\omega+\Sigma^A_{-Q})+\Gamma^R_Q\Gamma^A_{-Q}}\left(\begin{array}{cc}\Sigma^A_{-Q}+\omega & -\Gamma^R_Q\\ -\Gamma^A_{-Q} & \Sigma^R_{Q}-\omega
\end{array}\right),\nonumber
\end{eqnarray}
with the off-diagonal self-energies $\Gamma^R_Q$ and the Green's functions $g^R_Q=-i\langle\ann{\alpha}{Q}^c\cre{\alpha}{Q}^q\rangle, g^A_Q=-i\langle\ann{\alpha}{Q}^q\cre{\alpha}{Q}^c\rangle, h^R_Q=-i\langle \ann{\alpha}{Q}^q\ann{\alpha}{-Q}^c\rangle$ and $h^A_Q=-i\langle\cre{\alpha}{-Q}^q\cre{\alpha}{Q}^c\rangle$. 
The diagrammatic representation for the retarded and Keldysh Green's functions in Nambu space is depicted in Fig.~\ref{fig:OffGreens}.

\begin{figure}
  \includegraphics[width=8.6cm]{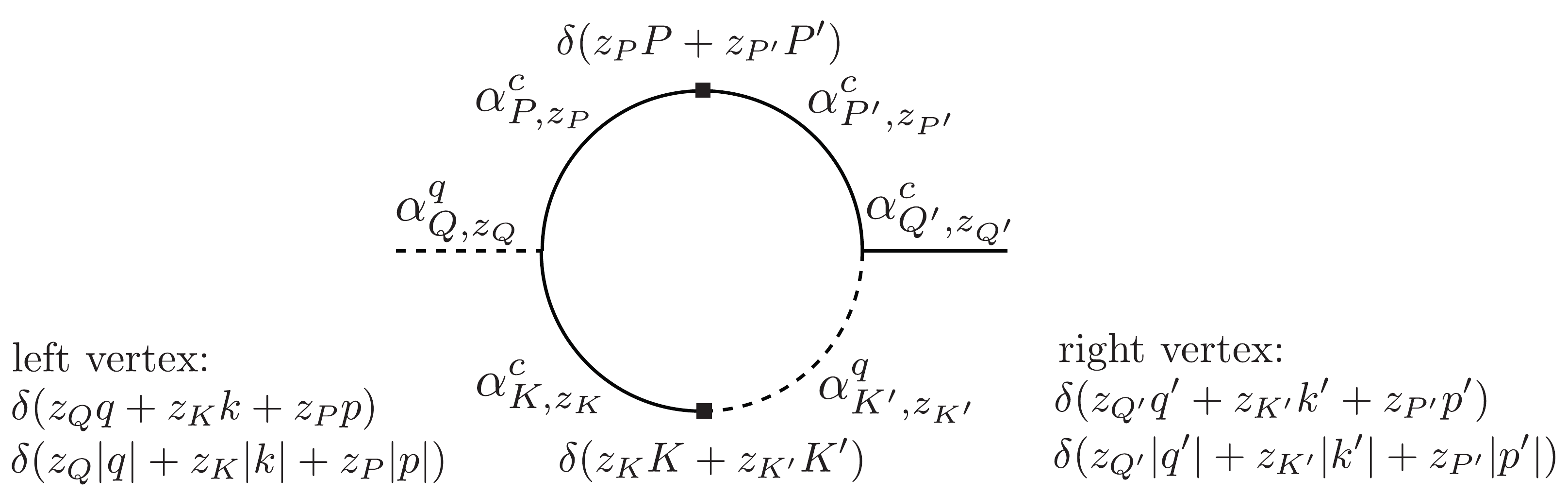}
  \caption{Diagrammatic representation of all possible one-loop diagrams contributing to the retarded sector of the self-energy. The additional index $z_{Q,P,K}=\pm 1$ represents ingoing lines ($\ann{\alpha}{Q,P,K}$ fields) for $z=1$ and outgoing lines ($\cre{\alpha}{Q,P,K}$ fields) for $z=-1$. The $\delta$-constraints stem from the momentum/frequency conservation in the Green's function and the momentum/frequency conservation in the two vertices. Two additional constraints, one for each vertex, are caused by the resonance condition. This results in six crucial constraints discussed in the main text.}
  \label{fig:OffDiagR}
\end{figure}

With this at hand, we can set up diagrammatic rules in Nambu space, as we do in the following. To this end, we start from the most general diagram contributing to the retarded sector of the Green's function, which is shown in Fig.~\ref{fig:OffDiagR}.
Here we replaced the fields $\ann{\alpha}{Q}\rightarrow\ann{\alpha}{Q,1}$ and $\cre{\alpha}{Q}\rightarrow \ann{\alpha}{Q,-1}$ in order to find a generalized representation of diagrams in Nambu space.
Exploiting frequency and momentum conservation in the Green's functions (diagonal and off-diagonal ones) and the vertices, together with the resonance condition at each vertex, we end up with the following relations
\begin{eqnarray}
z_QQ&=&z_{Q'}Q',\\
z_PP&=&z_{P'}P',\\
z_KK&=&z_{K'}K',\\
z_QQ&=&-z_KK-z_PP,\\
z_Q|q|+z_p|p|&=&-z_K|z_Qq+z_Pp|,\\
 \label{OffDiag12}z_Kz_{K'}+z_Qz_{Q'}&=&2z_Pz_{P'}.
\end{eqnarray}
Eq.~\eqref{OffDiag12} is a result of the two independent resonance conditions in Fig.~\ref{fig:OffDiagR} and reduces the number of diagrams for the retarded self-energy significantly. As a result of Eq.~\eqref{OffDiag12}, the product of all the corresponding $z$-factors must be identical. For the diagonal self-energy, where $z_Qz_{Q'}=-1$, this means that $z_Pz_{P'}=z_Kz_{K'}=-1$ and consequently only loops consisting of two diagonal Green's functions contribute to the diagonal self-energy. Consequently, the diagrammatic representation of the diagonal self-energy is still given by the loops shown in Fig.~\ref{fig:Diagrams} even in the presence of off-diagonal Green's functions, i.e.
\begin{eqnarray}
-i\Sigma^R_Q&=&\frac{v_0^2}{4\pi^2}\int_P|pq|\left(|q-p|g^K_Pg^R_{Q-P}+|p+q|g^K_{Q+P}g^A_{P}\right.\nonumber\\
&&\phantom{\frac{v_0^2}{4\pi^2}\int_P|pq|}\left.+|p+q|g^K_{P}g^R_{P+Q}\right).\label{OffDiag13}
\end{eqnarray}
In order to obtain the off-diagonal self-energy $\Gamma^R_Q$, one has to flip the sign of $z_{Q'}\rightarrow 1$, such that $z_Qz_{Q'}=z_Pz_{P'}=z_Kz_{K'}=1$ in the corresponding loops. This means that the diagrams contain only off-diagonal  terms. $\Gamma^R_Q$ is then obtained by the diagrams in Fig.~\ref{fig:Diagrams}, but with all arrows from the right vertex flipped. We thus obtain
\begin{eqnarray}
-i\Gamma^R_Q&=&\frac{v_0^2}{4\pi^2}\int_P|pq|\left(|q-p|h^K_Ph^R_{Q-P}+|p+q|h^K_{Q+P}h^A_{P}\right.\nonumber\\
&&\phantom{\frac{v_0^2}{4\pi^2}\int_P|pq|}\left.-|p+q|\left(h^K_{P}\right)^*h^R_{P+Q}\right).\label{OffDiag14}
\end{eqnarray}
The diagonal self-energy $\Sigma^R$ diverges when the integral \eqref{OffDiag13} is performed with the bare Green's functions. This hints that a non-trivial self-energy is generated on the diagonal to regulate the integral, which can be computed in self-consistent Born approximation as explained in previous sections. On the other hand, for off-diagonal Green's functions $h^R=h^A=0$ (e.g. for the bare off-diagonal values), the off-diagonal self-energy is zero, as visible from Eq.~\eqref{OffDiag14}. Consequently, off-diagonal self-energies are not generated in the retarded sector even in self-consistent Born approximation and we have
\eq{OffDiag15}{
\Gamma^R_Q=0.} 
In the absence of off-diagonal self-energies in the retarded sector, we can directly apply the quasi-particle approximation discussed in the previous sections and evaluate the self-energies and distribution function on-shell. Consequently, the intermediate kinetic equation is
\eq{OffDiag16}{
\partial_{\tau}F_{q,\omega=0,\tau}=-2\sigma^R_{q,\tau}F_{q,\omega=0,\tau}+i\sigma_z\Sigma^K_{q,\omega=0,\tau}\sigma_z}
with the scalar, on-shell self-energy $\sigma^R_{q,\tau}$ as discussed in Sec. \ref{sec:OnShell}. 

\begin{figure}[h]
  \includegraphics[width=8.6cm]{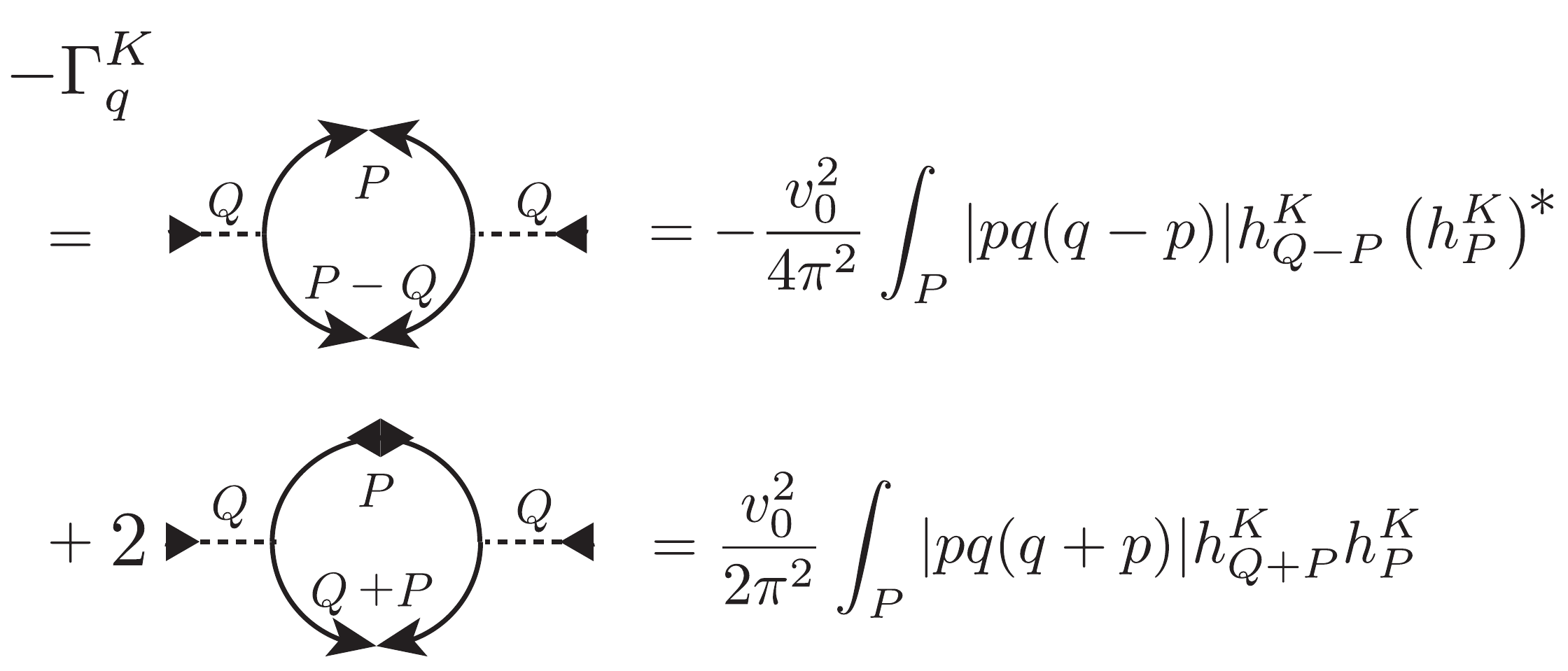}
  \caption{Diagrammatic representation of the off-diagonal Keldysh self-energy $\Gamma^K_q$. Compared to the diagonal component, for the vertex of the left ingoing and outgoing lines have been replaced. The diagrams containing only retarded/advanced Green's functions, which contributed to $\sigma^K_q$, are absent because of the absence of off-diagonal retarded and advanced Green's functions.}
  \label{fig:OffDiagK}
\end{figure}

For the on-shell Keldysh self-energy,
\eq{OffDiag17}{
\Sigma^K_{q,\omega=0,\tau}=-2i\left(\begin{array}{cc} \sigma^K_{q,\tau}& \Gamma^K_{q,\tau}\\ \Gamma^K_{q,\tau} & \sigma^K_{q,\tau}\end{array}\right)
} the diagrammatic rules from the previous section do not have to be modified. This immediately yields the diagonal, on-shell Keldysh self-energy $\sigma^K_{q,\tau}$ according to Fig.~\ref{fig:DiagramsK}. For the off-diagonal, on-shell Keldysh self-energy $\Gamma^K_q$ the corresponding diagrams are obtained by reversing the arrows associated to the vertices on the right in Fig.~\ref{fig:DiagramsK}, resulting in the diagrammatic representation of $\Gamma^K_q$ shown in Fig.~\ref{fig:OffDiagK}.
The Keldysh Green's functions are obtained via the parametrization used in Eq.~\eqref{OffDiag8}, which yields for the off-diagonal elements
\eq{OffDiag18}{
h^K_{q,\omega,\tau}=\frac{16\pi^2i\sigma^R_{q,\tau}m_{q,\tau}}{\left(\omega-u|q|\right)^2+\left(\sigma^R_{q,\tau}\right)^2}.}
Evaluation of the off-diagonal diagrams and insertion into the kinetic equation completes the set of equations for a system including anomalous phonon densities. 

Finally, the off-diagonal retarded/advanced self-energies are exactly zero and the kinetic equation for the off-diagonal terms is
\begin{eqnarray}
\partial_{\tilde{\tau}}m_q&=&\int_{0<p<q}\frac{2pq(q-p)}{\tilde{\sigma}^R_q+\tilde{\sigma}^R_p+\tilde{\sigma}^R_{q-p}}\left(m_pm_{q-p}-m_q\left(1+n_p+n_{q-p}\right)\right)\ \ \nonumber\\
&+&\int_{0<p}\frac{4pq(q+p)}{\tilde{\sigma}^R_q+\tilde{\sigma}^R_p+\tilde{\sigma}^R_{q+p}}\left(n_{p+q}m_q+m_pm_{p+q}-m_qn_p\right).\label{OffDiag19}
\end{eqnarray}
Here, we have again used the transformed basis to eliminate the factor $v_0^2$ in front of the integrals. The time evolution for the off-diagonal terms depends on both the diagonal and off-diagonal densities, and the stationary solution of this equation is $m_q=0$ due to the uncompensated spontaneous decay term in the first integral. Together with Eqs. \eqref{SelfEn15} and \eqref{Kinetic5}, Eq.~\eqref{OffDiag19} represents the complete set of equations determining the time evolution of the phonon density and the self-energies of an interacting Luttinger Liquid for a system that has been initialized in an out-of-equilibrium state $n_q\neq n\sub{B}(u|q|), m_q\neq0$.

\section{Relaxation of an excited thermal state}\label{sec:Sec6}
In this section, we will analyze the relaxation dynamics of a nearly thermal initial state and compare it to the analytical results from the previous sections. We consider an initial state with the densities
\begin{eqnarray}
n_{q,\tau=0}&=&n\sub{B}(T\sub{i}=2u|q_0|)+\delta_q\ \ \nonumber\\
&=&\frac{1}{e^{0.5\left|\frac{q}{q_0}\right|}-1}+\alpha_1 e^{-\frac{(q-q_0)^2}{2\alpha_2^2}},\label{Sim1}\\
m_{q,\tau=0}&=&\delta m_{q,\tau=0}=2\alpha_1 e^{-\frac{(q-q_0)^2}{2\alpha_2^2}}.\label{Sim2}
\end{eqnarray}
A state of this form can be created by perturbing a thermal state with initial temperature $T\sub{i}=0.5u|q_0|$ in coupling the operator $\propto \partial_x\phi$ to a classical field with momentum $q_0$, i.e. in a microscopic fermionic or bosonic model by a small density modulation with momentum $q_0$\cite{giamarchi04}. For a specific simulation, we chose the parameters $\alpha_1=0.2, \alpha_2=4q_0$ and express momentum in units of $q_0$.
\begin{figure}
  \includegraphics[width=8.6cm]{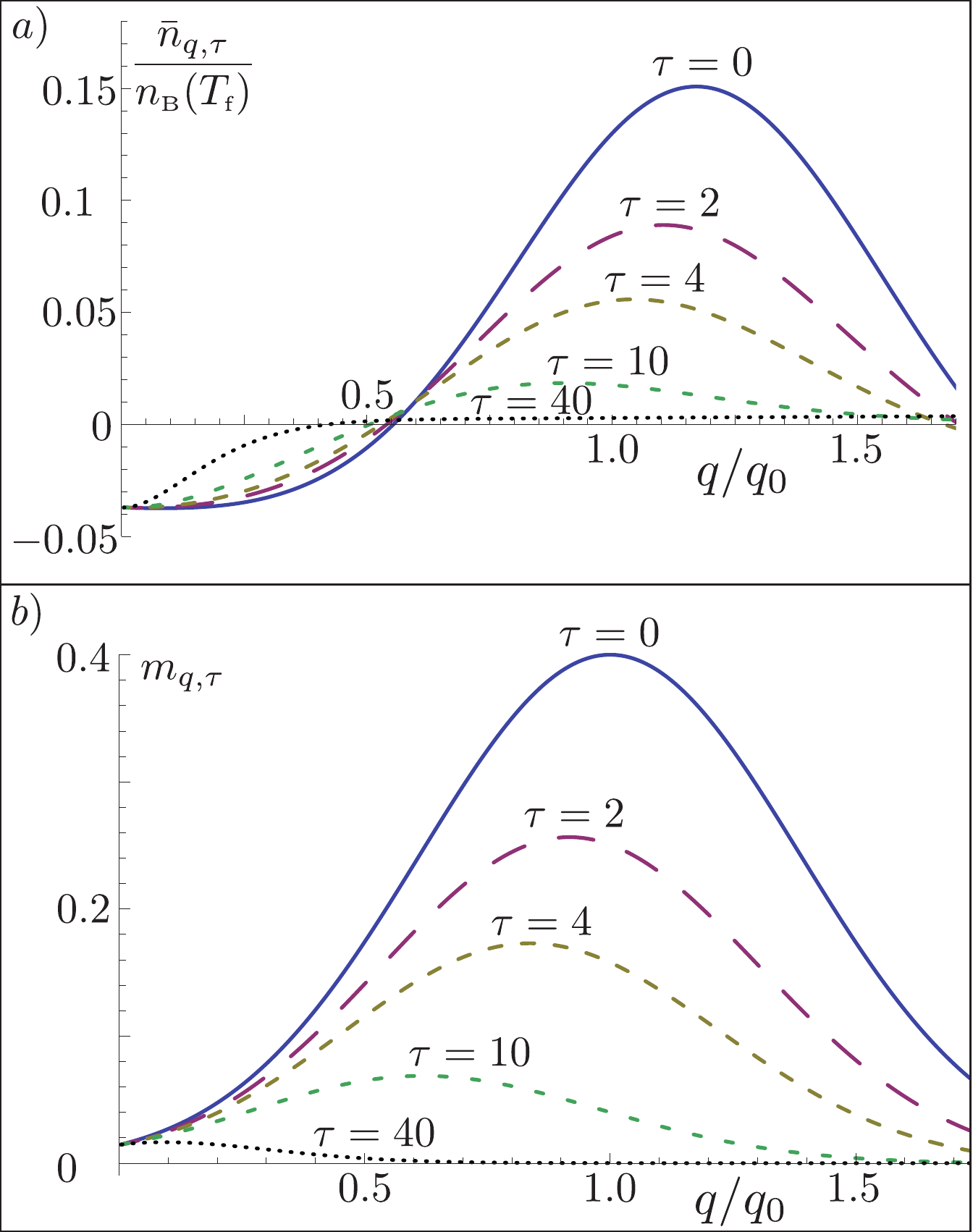}
  \caption{Time evolved phonon density $n_{q,\tau}$ after initializing the system in a state described by Eqs.~\eqref{Sim1}, \eqref{Sim2}. The time $\tau$ is expressed in units of $\frac{1}{v_0q_0^2}$. a) The deviation $\bar{n}_{q,\tau}$ of $n_{q,\tau}$ from the final thermal density $n\sub{B}(T\sub{f})$ divided by $n\sub{B}(T\sub{f})$. $\bar{n}_{q,\tau}$ decays to zero with a momentum dependent rate $\gamma_q\approx -1.1 \sigma^R_q$. For smaller momenta, this rate decreases and finally becomes zero for $q\rightarrow0$. b) Anomalous density $m_{q,\tau}$ for different times $\tau$. The pinning of $m_{q=0,\tau}$ is the consequence of global particle number and current conservation of the present model.}
  \label{fig:Decays}
\end{figure}
The perturbation increases the energy 
\eq{Sim3}{
E(\tau)=\int_q u|q| n_{q,\tau}}
of the system. As a result, the final state in the limit $\tau\rightarrow \infty$ will be a thermal state with increased temperature $T\sub{f}>T\sub{i}$. Since the kinetics is energy conserving, $E(\tau)=E(\tau=0)$ for all $\tau>0$, the temperature can be determined according to
\eq{Sim4}{
E=\int_q \frac{u|q|}{e^{u|q|/T\sub{f}}-1}=\frac{T\sub{f}^2\pi^2}{6u}}
from the system energy. In the present case, this leads to $T\sub{f}=0.52 u|q_0|$ and a final state of the system
\eq{Sim5}{
\lim_{\tau\rightarrow\infty}n_{q,\tau}=n\sub{B}(T\sub{f}).}
The quantity of interest is the deviation of the time-dependent phonon density from the final phonon density in the limit ${\tau\rightarrow\infty}$, 
\eq{Sim6}{
\delta n_{q,\tau}\equiv n_{q,\tau}-n\sub{B}(T\sub{f}).}
In Fig.~\ref{fig:Decays}, we show $\delta n_{q,\tau}$ and $m_{q,\tau}$ for different time steps $\tau_{l}$, and we see in which way both quantities decay to zero momentum as a function of time and momentum.
\begin{figure}
  \includegraphics[width=8.6cm]{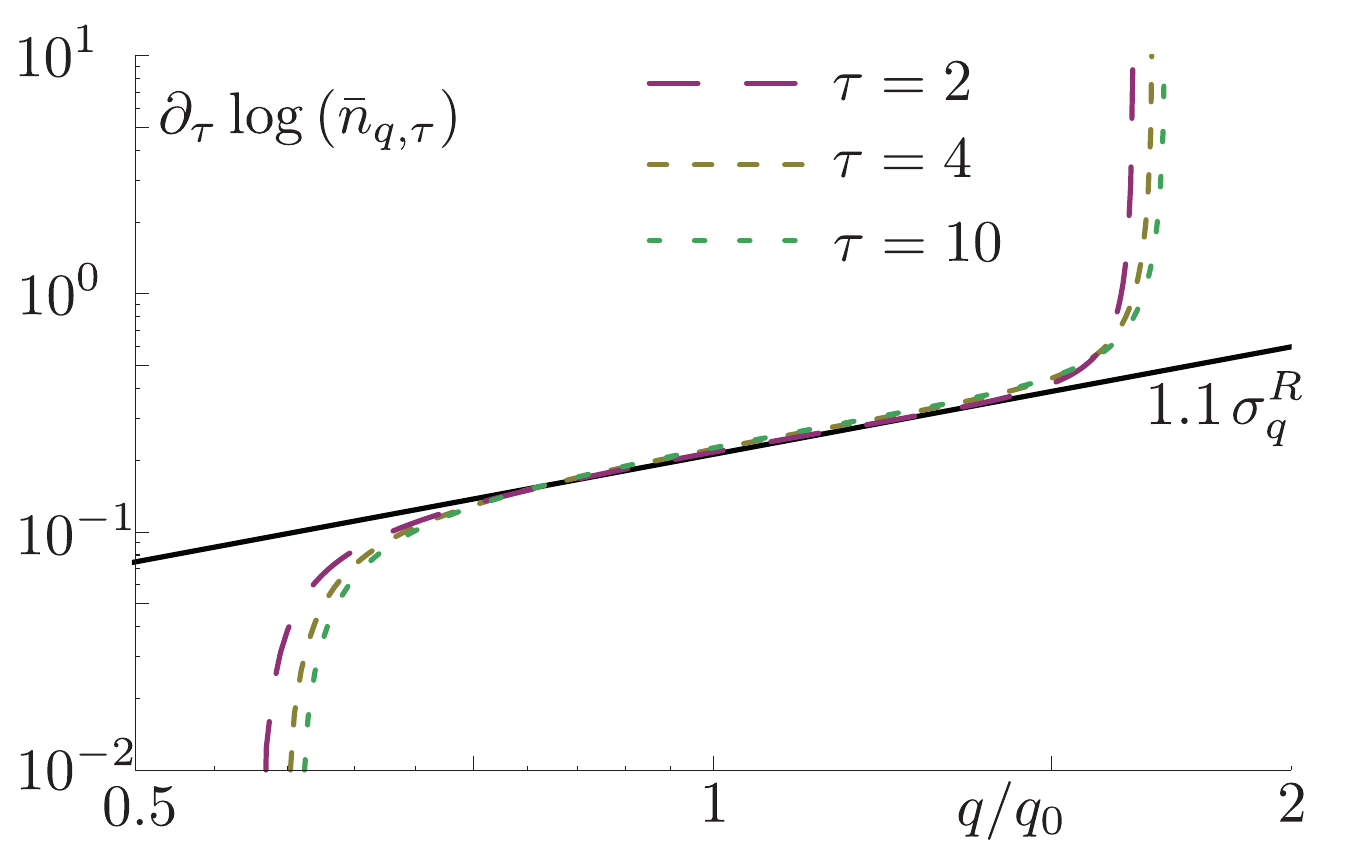}
  \caption{Comparison of the decay rate $\gamma_q=-\partial_{\tau}\log\left(\bar{n}_{q,\tau}\right)$ as a function of momentum for different times $\tau$ with the analytically estimated decay factor $\gamma_q^A=2.2\sigma^R_q$. The time is given in units of $\frac{1}{v_0q_0^2}$.  The exact decay coincides very well with the analytical prediction. Only for  $\bar{n}_{q,\tau}\approx 0$, small numerical errors leading to $\partial_{\tau}\bar{n}_q\neq0$ lead to a divergence of the logarithm.}
  \label{fig:DecayRate}
\end{figure}

The evolution of this excited state to its new equilibrium proceeds in two different stages\cite{Lin13}: For short times, the set of excited modes is much more strongly occupied than all other modes, and the dominant effect is the decay of these modes into the continuum of non-excited ones. In this step, there is no back-action from the continuum, which acts as a bath for the excited states. As a consequence, the dynamics is described by the corresponding equilibrium linear response.
According to Eq.~\eqref{Kinetic11}, $\delta n_q$ then follows the effective equation of motion
\eq{Sim7}{
\partial_{\tau}\log\left(\delta n_{q,\tau}\right)= -1.1\sigma^R_{q}.}
As a result, for short times $\delta n_{q,\tau}$ decays exponentially in time with a momentum dependent decay time $\tau\sub{dec}(q)=\left(1.1\sigma^R_q\right)^{-1}$.

In Fig.~\ref{fig:DecayRate}, we compare the numerical value of $\partial_{\tau}\log\left(\delta n_{q,\tau}\right)$
with the analytical estimate\\ ${\partial_{\tau}\log\left(\delta n_{q,\tau}\right)\approx -1.1\sigma^R_{q,\tau}\approx 0.87 \sqrt{\frac{2\pi T\sub{f}}{u}}q^{\frac{3}{2}}}$ and find a good agreement in the momentum region where $\delta n_q$ deviates from zero.
\begin{figure}
  \includegraphics[width=8.6cm]{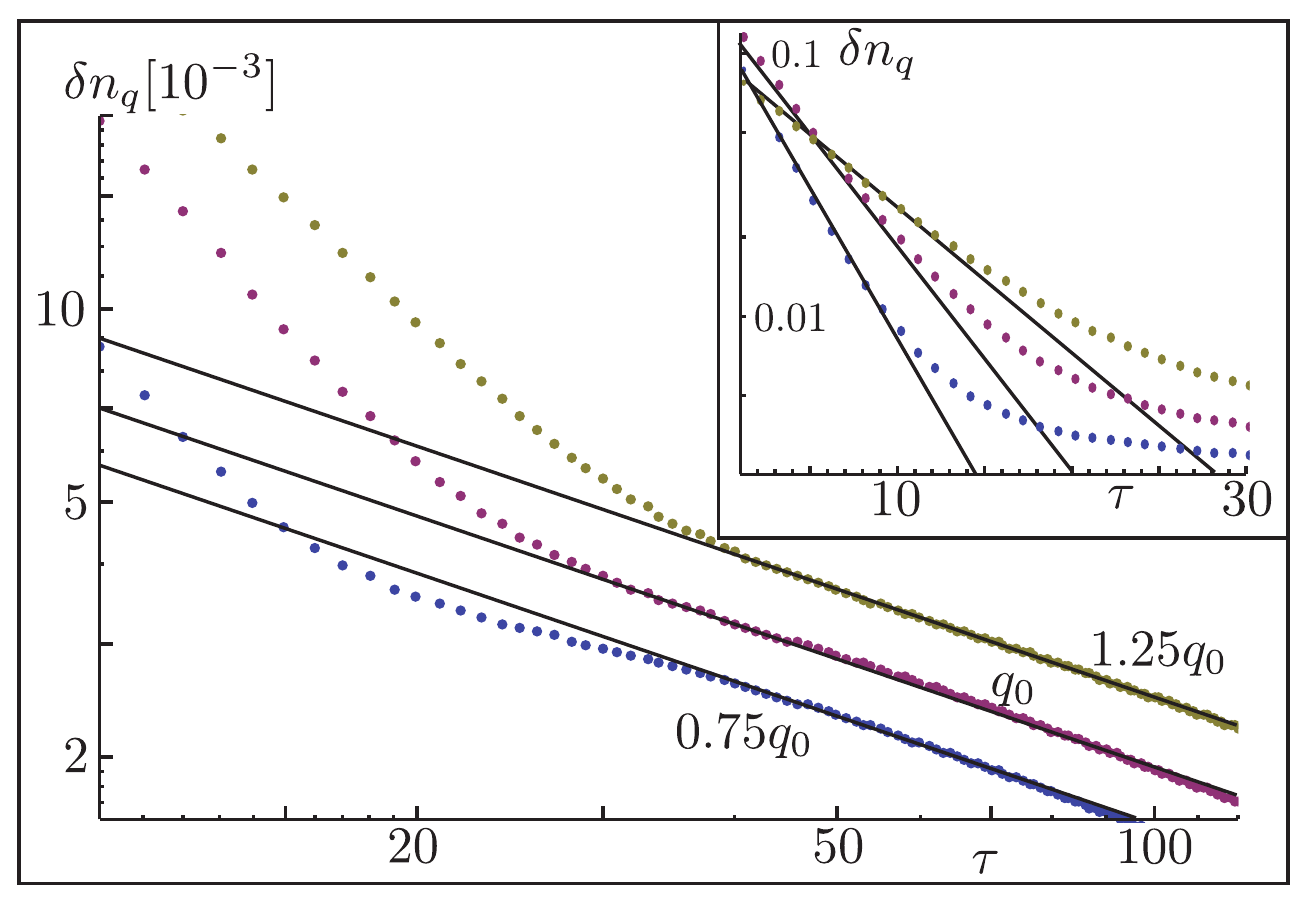}
  \caption{Decay of the deviation $\delta n_q(\tau)$ of the phonon density to its corresponding thermal value for fixed momenta $(q_1,q_2,q_3)=(0.75,1,1.25)q_0$. a) Log-Log plot of the deviation $\delta n_q(\tau)$ (blue dots: simulations; line: algebraic fit): For long times, the decay is described by a power law in time, i.e. $\delta n_q(\tau)\sim \tau^{-\beta}$ with an exponent $\beta\approx 0.58$. b) Semi-logarithmic plot of the deviation $\delta n_q(\tau)$: For short times, the decay is described by an exponential, $\delta n_q(\tau)\sim e^{-1.1\sigma^R_{q_0}\tau}$, where $\sigma^R$ is the corresponding thermal self-energy (blue dots: numerical simulation; red line: exponential fit, obtained from the self-energy). }
  \label{fig:ExpoVSAlg}
\end{figure}

For larger times instead, many modes deviate only very weakly from the thermal occupation, but the system has still not found its equilibrium. In this case, back-action from the mode continuum can no longer be ignored. More precisely, due to energy and momentum conservation, this dynamics, which now is dominated by in- and out-scattering events on an equal footing, is very slow. It is no longer described by an equilibrium response theory, which is determined solely by the retarded self-energy. Instead the presence of dynamical slow modes is revealed, which are not captured by a simple exponential decay but have to be implemented as additional modes due to symmetries\cite{Lux13} and favor an algebraic decay in time, i.e. \\ ${\delta n_q(\tau)\sim \tau^{-\beta}}$ with some exponent $\beta$. In Fig.~\ref{fig:ExpoVSAlg}, the two different stages of the time evolution are visible in the numerical simulation of the quasi-particle occupation. The algebraic decay for long times, and the exponential decay for shorter times, are clearly distinguished. In order to estimate the algebraic exponent, we do not linearize the kinetic equation but instead take Eq.~\eqref{Kinetic5} and insert on the right hand side the solution for the phonon occupations $n_q(\tau)-n\sub{B}(T\sub{f})\sim e^{-1.1\sigma^R_q\tau}$. Due to the scaling of the thermal self-energy $\sigma_q^R\sim q^{\frac{3}{2}}$, this yields the estimate $\beta=\frac{2}{3}$.  In the numerical simulations presented in Fig.~\ref{fig:ExpoVSAlg}, we find for the algebraic exponent $\beta\approx 0.58$.

As mentioned already, this algebraic decay is a consequence of energy and momentum conservation in the dynamics, which leads to additional slow modes in the time evolution. The separation of the relaxation into two different time regimes, with first exponential decay according to bare phonon decay and then algebraic decay due to energy conservation, has already been found in a recent work using an equilibrium formalism \cite{Lin13}. In this work, energy conservation is ensured by an effective, time-dependent temperature, which can, however, not account for local energy fluctuations. On the other hand, in the formalism presented in this manuscript, both spatial and temporal energy fluctuations are naturally incorporated in the kinetic equation.
The algebraic decay of the phonon occupations can be explained in terms of dynamical slow modes, resulting from conservation laws, i.e. symmetries, in the system. The analytical exponent $\beta=\frac{2}{3}$ would belong to a system with exact momentum and energy conservation with a classical distribution function $n_q\sim \frac{1}{|q|}$. The deviation of the numerical result $\beta\approx0.58$ from the analytical one might be a result either of subleading corrections that will vanish on even larger time scales\cite{Lux13} or the fact that the system is not described by a classical distribution function $n_q\sim\frac{1}{|q|}$ for the complete momentum range but by a Bose distribution, which seriously differs from the classical one at intermediate and large momenta.

\section{Dyson-Schwinger Equations and Vertex Corrections}\label{sec:DS}

\begin{figure}[t!]
  \includegraphics[width=7cm]{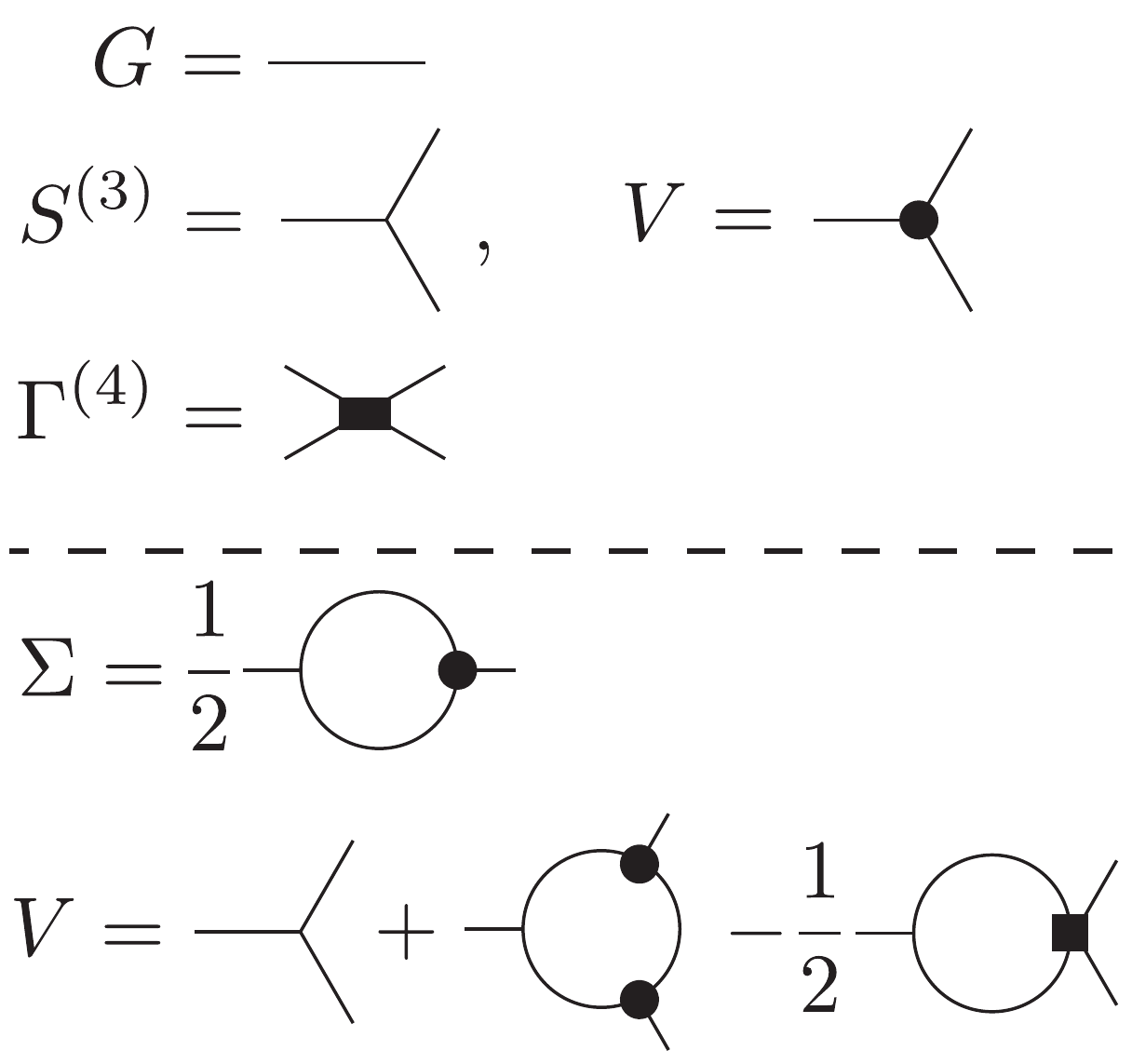}
  \caption{Schematic diagrammatic representation of the self-energy $\Sigma$ and the full three-point vertex $V$ in terms of the full Green's function $G$, bare three-point vertex $\tilde{V}$, and the full four-point vertex $\Gamma^{(4)}$.}
  \label{fig:DiagsDS}
\end{figure}
In this section we apply Dyson-Schwinger equations \cite{Amit,peskinschroeder} to the interacting Luttinger Liquid and determine the self-energy and three-point vertex as a function of the phonon distribution. We show that the vertex correction is always real and will be exactly zero for a zero temperature state. More generally it leads to a modification $v_0\rightarrow v_0\left(1+\mathcal{I}\right)$, where $\mathcal{I}$ is a small $(1\gg\mathcal{I}$) dimensionless function with weak momentum dependence, whose precise form is determined by the time-dependent phonon density. Based on these findings, we conjecture that also the corrections to the four-point and higher order vertices, which do not occur in the microscopic action, are small.
As a result, the kinetic equation and the equation for the self-energies in self-consistent Born-approximation are modified according to the vertex correction, and we derive a coupled but closed set of equations. 

In general, Dyson-Schwinger equations (DSE) represent an exact hierarchy for one-particle irreducible (1PI) correlation functions, generated by the effective action functional $\Gamma[\ann{a}{\alpha},\ann{a}{\beta}]$ \cite{Amit,peskinschroeder,zinnjustinbook}. It relates the full vertices (e.g. the Green's function, which is the inverse two-point vertex) to their bare, microscopic counterparts, which form the microscopic action $S$. The hierarchy built up by the DSE is in general infinite and therefore lacking an exact solution. The main goal is then to find a physically reasonable truncation of the effective action, for which the main physical effects are captured and in which the DSE can be solved.

The effective action, as the generator of 1PI correlation functions, can be expandend according to\cite{Amit,peskinschroeder,zinnjustinbook}
\eq{DS3}{
\Gamma[\ann{a}{\alpha},\ann{a}{\beta}]=\sum_{n=2}^{\infty}\frac{1}{n!}\Gamma^{(n)}_{\alpha_1,...\alpha_n}\ann{a}{\alpha_1}...\ann{a}{\alpha_n},}
where $\Gamma^{(n)}_{\alpha_1,...\alpha_n}$ is the $n$-th order functional derivative of the effective action,
\eq{DS4}{
\Gamma^{(n)}_{a_1,...a_n}=\left.\frac{\delta^n\Gamma}{\delta a_n...\delta a_1}\right|_{\ann{a}{a_i}=0},}
i.e. the full $n$-point vertex function, and $\ann{a}{\alpha}$ are the fields with collective index $\alpha=(\omega,q,\tau,c/q)$.
The inverse Green's function and the full three-point vertex are the second and third order functional derivatives, respectively. In terms of formulas we have
\eq{DS5}{
G^{-1}=G_0^{-1}-\Sigma=\Gamma^{(2)}\ \mbox{ and }\ V=\Gamma^{(3)}.}
%The effective action is determined by the Dyson-Schwinger master equation\eq{DS6}{\frac{\delta\Gamma}{\delta\ann{a}{\alpha}}=\frac{\delta S}{\delta\ann{a}{\alpha}}\left[\ann{a}{\alpha}+i\left(\frac{\delta^2\Gamma}{\delta a^2}\right)_{\alpha,\beta}^{-1}\frac{\delta}{\delta\ann{a}{\beta}}\right]\cdot 1,}where $S$ is the bare action of the interacting Luttinger Liquid given in Eq.~\eqref{Model18} and the fields have been replaced by the argument in brackets. As a result we find\eq{DS7}{\left(\frac{\delta\Gamma}{\delta a}\right)_{\alpha}=\left(G_0^{-1}\right)_{\alpha\beta}\ann{a}{\beta}+\frac{1}{2}S^{(3)}_{\alpha\beta\gamma}\left(\ann{a}{\beta}\ann{a}{\gamma}+i\left(\frac{\delta^2\Gamma}{\delta a^2}\right)^{-1}_{\beta\gamma}\right)}for the first functional derivative of the effective action and \eq{DS8}{\left(\frac{\delta^2\Gamma}{\delta a^2}\right)_{\alpha\beta}=\left(G_0^{-1}\right)_{\alpha\beta}+S^{(3)}_{\alpha\gamma\delta}\left(\ann{a}{\gamma}\delta_{\delta\beta}-\frac{i}{2}\left(\frac{\delta^2\Gamma}{\delta a^2}\right)^{-1}_{\gamma\mu}\left(\frac{\delta^3\Gamma}{\delta a^3}\right)_{\mu\beta\nu}\left(\frac{\delta^2\Gamma}{\delta a^2}\right)^{-1}_{\nu\delta}\right)}for the second derivative. In order to determine the full Green's function, according to \eqref{DS5}, the fields $a$ are set to zero in Eq.~\eqref{DS8}, leading to

In the present case, the DSE relate the $n$-point vertex to the $n+1$ point vertex of the theory. More specifically, we obtain for the full Green's function
\eq{DS9}{
\left(G^{-1}\right)_{\alpha\beta}=\left(G^{-1}_0\right)_{\alpha\beta}-\tfrac{i}{2}S^{(3)}_{\alpha\gamma\delta}G_{\gamma\mu}V_{\mu\beta\nu}G_{\nu\delta},}
and for the full three-point vertex
\begin{eqnarray}
V_{\alpha\beta\gamma}&=&S^{(3)}_{\alpha\beta\gamma}+iS^{(3)}_{\alpha\delta\nu}G_{\delta\mu}V_{\mu\beta\eta}
G_{\eta\sigma}V_{\sigma\gamma\epsilon}
G_{\epsilon\nu}\nonumber\\
&&-\tfrac{i}{2}S^{(3)}_{\alpha\delta\nu}G_{\delta\mu}\Gamma^{(4)}_{\mu\beta\gamma\eta}
G_{\eta\nu}\label{DSc9}.
\end{eqnarray}
Here, $S^{(3)}$ is the bare (microscopic) three-point vertex. A schematic diagrammatic representation of Eqs. \eqref{DS5}, \eqref{DS9} and \eqref{DSc9} is depicted in Fig.~\ref{fig:DiagsDS}.

We will show in the following that there are corrections to the bare three-point vertex, with the same scaling dimension as the bare three-point vertex itself. It is determined by a dimensionless function $\mathcal{I}$ whose precise form is dictated by the time dependent phonon density $n_{q,\tau}$. 
This is in contrast to the two-point vertex, i.e. the inverse Green's function, where the correction due to the cubic vertex is subleading but introduces an imaginary part and therefore is of great qualitative importance. Here the correction is purely real, as the bare vertex itself, but on the other hand not subleading. It therefore can not be discarded directly without further discussion.

\begin{figure}
  \includegraphics[width=8.6cm]{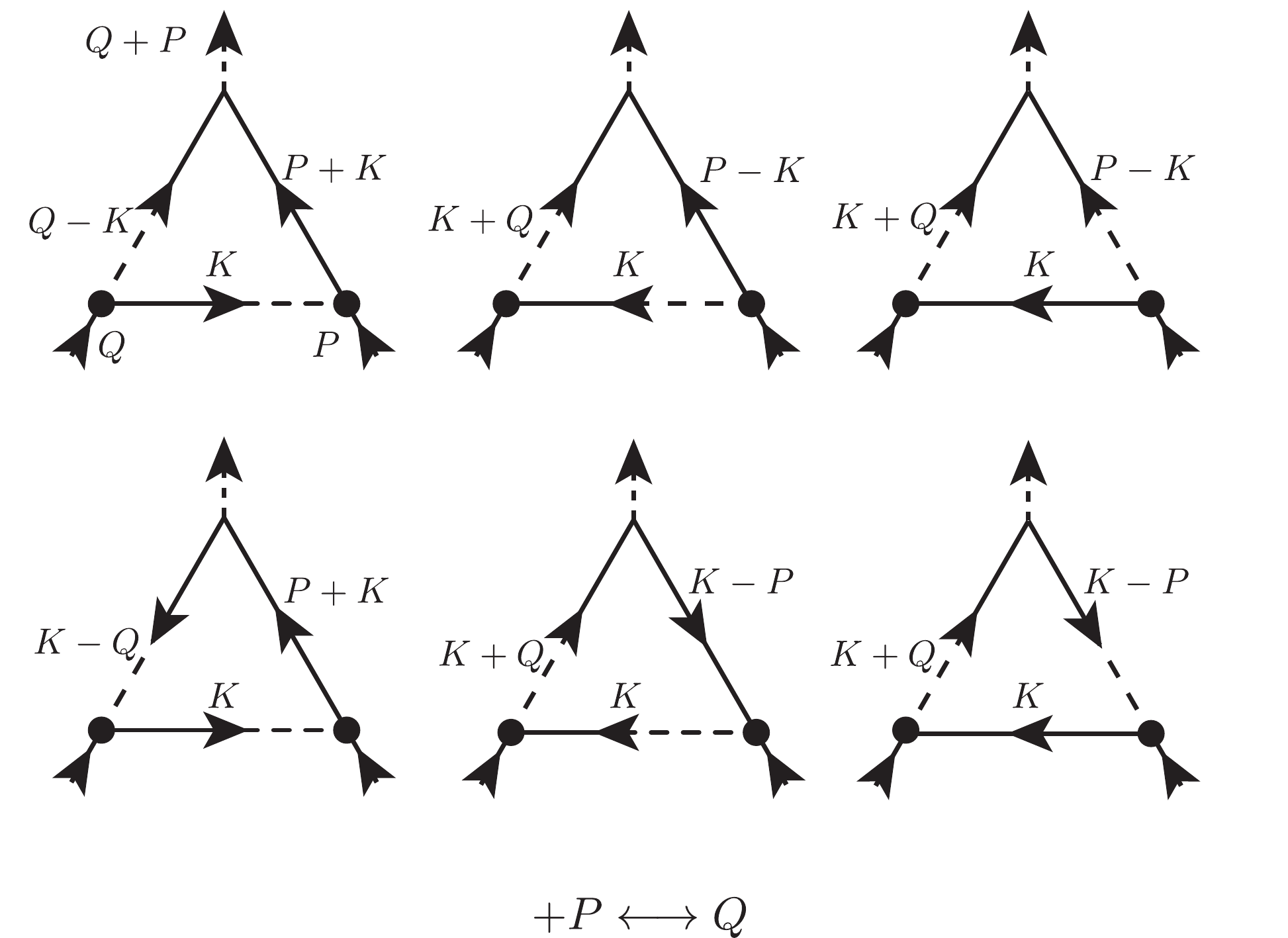}
  \caption{Diagrams contributing to the vertex correction of the three-point vertex $V_{q,p,p+q}$ illustrated for the particular example of an outgoing quantum field and two incoming classical fields, i.e. $\propto\cre{a}{Q+P}^q\ann{a}{Q}^c\ann{a}{P}^c$. In total there are twelve distinct diagrams contributing to the vertex correction, six are depicted above and six further can be found by interchanging the ingoing momenta $P\leftrightarrow Q$.}
  \label{fig:VertexCorr}
\end{figure}

For the $\cre{a}{Q+P}^q\ann{a}{P}^c\ann{a}{Q}^c$ term, the vertex correction $\delta V=V-S^{(3)}$ is illustrated in a diagrammatic representation in Fig.~\ref{fig:VertexCorr}. It is equivalent to all other vertices that incorporate a single quantum and two classical fields. The lowest order contribution, incorporating only bare vertices, reads (for $q,p>0$)

\begin{widetext}
\eq{DSc10}{
\delta V_{q,p,p+q}^{ccq}=\frac{v_0^3}{\sqrt{8}}\sqrt{|qp(p+q)|}\int_{k>0}\left\{ \frac{k(q+k)(k-p)}{\sigma^R_{k+q}+\sigma^R_{k-p}}\left[\frac{n_{k-p}-n_k}{\sigma^R_k+\sigma^R_{k-p}}+\frac{n_{q+k}-n_k}{\sigma^R_k+\sigma^R_{k+q}}
\right]+\frac{k(p+k)(k-q)}{\sigma^R_{k+p}+\sigma^R_{k-q}}\left[\frac{n_{k-q}-n_k}{\sigma^R_{k-q}+\sigma^R_k}+\frac{n_{p+k}-n_k}{\sigma^R_{k+p}+\sigma^R_k}\right]\right\}.
}
\end{widetext}

Here $\sigma^R_{k}=-\mbox{Im}(\Sigma^R_{K})>0$ are the on-shell self-energies obtained by the DSE in Fig.~\ref{fig:DiagsDS}.  The vertex correction due to cubic vertices is identical to zero for a zero temperature system ($n_k=0$ for all momenta). This is an exact statement for the interacting Luttinger Liquid and can be seen on the level of the diagrams in Fig.~\ref{fig:VertexCorr}. For $T=0$, $G^K=\mbox{sgn}(\omega)\left(G^R-G^A\right)$ and the individual diagrams cancel each other due to the pole structure of $G^R$ and $G^A$. In general, for a constant quasi-particle distribution function, the cubic vertex correction is exactly zero in arbitrary dimensions \cite{forsternelson76}.

In order to find a compact expression for the vertex correction, we replace $\sigma^R\rightarrow v_0\tilde{\sigma}^R$ and compare the integrand in Eq.~\eqref{DSc10} with the expression for the self-energy $\tilde{\sigma}^R$ in Eq.~\eqref{SelfEn18}. We immediately see, that the vertex correction is linear in $v_0$ and the integral has scaling dimension zero\footnote{This holds true for an arbitrary phonon density $n_q$ since the dimensions cancel exactly, independent of the form of $n_q$.}. 
In perturbation theory, this yields the vertex
\begin{eqnarray}
V_{q,p,q+p}^{(\mbox{\tiny 1st})}&=&v_0\sqrt{|pq(p+q)|}\left(1+\mathcal{I}_0\left(\frac{p}{q},n\right)\right)\nonumber\\
&=&S^{(3)}_{q,p,p+q}\left(1+\mathcal{I}_0\left(\frac{p}{q},n\right)\right).\label{DSc11}
\end{eqnarray}
Here, $\mathcal{I}_0\left(\frac{p}{q},n\right)$ is a dimensionless function of the ratio $p/q$ and the phonon density $n$, which is determined by the integral in Eq.~\eqref{DSc10}. The scaling behavior of the one-loop vertex correction suggests the parametrization of the full vertex according to
\begin{eqnarray}
V_{q,p,q+p}&=&v_0\sqrt{|pq(p+q)|}\left(1+\mathcal{I}\left(\frac{p}{q},n\right)\right),\label{DScEx}
\end{eqnarray}
where the functional $\mathcal{I}$ encodes the full vertex correction. According to the DSE in Fig.~\ref{fig:DiagsDS}, it is determined via

\begin{widetext}
\begin{eqnarray}
\mathcal{I}(x,n)&=&\frac{1}{\sqrt{8}}\int_{{\tilde{k}}>0}\left\{ \frac{{\tilde{k}}(1+{\tilde{k}})({\tilde{k}}-x)\left(1+\mathcal{I}\left(\frac{{\tilde{k}}}{|1-{\tilde{k}}|},n\right)\right)\left(1+\mathcal{I}\left(\frac{{\tilde{k}}}{x},n\right)\right)}{\tilde{\sigma}^R_{{\tilde{k}}+1}+\tilde{\sigma}^R_{{\tilde{k}}-x}}\left[\frac{n_{{\tilde{k}}-x}-n_{\tilde{k}}}{\tilde{\sigma}^R_{\tilde{k}}+\tilde{\sigma}^R_{{\tilde{k}}-x}}+\frac{n_{1+{\tilde{k}}}-n_{\tilde{k}}}{\tilde{\sigma}^R_{\tilde{k}}+\tilde{\sigma}^R_{{\tilde{k}}+1}}
\right]\right.\nonumber\\
&&\left.+\frac{{\tilde{k}}(x+{\tilde{k}})({\tilde{k}}-1)\left(1+\mathcal{I}\left(\frac{x}{|{\tilde{k}}-x|},n\right)\right)\left(1+\mathcal{I}\left({\tilde{k}},n\right)\right)}{\tilde{\sigma}^R_{{\tilde{k}}+x}+\tilde{\sigma}^R_{{\tilde{k}}-1}}\left[\frac{n_{{\tilde{k}}-1}-n_{\tilde{k}}}{\tilde{\sigma}^R_{{\tilde{k}}-1}+\tilde{\sigma}^R_{\tilde{k}}}+\frac{n_{x+{\tilde{k}}}-n_{\tilde{k}}}{\tilde{\sigma}^R_{{\tilde{k}}+x}+\tilde{\sigma}^R_{\tilde{k}}}\right]\right\}.\label{DSc12}
\end{eqnarray}
\end{widetext}

where $\tilde{k}=\frac{k}{q}$ and the integral is dimensionless.
In Eq.~\eqref{DSc12}, we have already exploited the fact that the ingoing momenta of a vertex can be exchanged without modifying the vertex itself, and consequently the integral is invariant under $q\leftrightarrow p$. This is equivalent to $\mathcal{I}\left(x,n\right)=\mathcal{I}\left(\frac{1}{x},n\right)$. The self-energy in the DSE approach is
\begin{eqnarray}
\tilde{\sigma}^R_q&=&\int_{0<p}\left(\frac{\partial_{\tilde{\tau}}n_p}{\tilde{\sigma}^R_p}+2n_p+1\right)\label{DSc13}\\
&&\left\{\left[1+\mathcal{I}\left(\tfrac{p}{|p-q|}\right)\right]\frac{qp(q-p)}{\tilde{\sigma}^R_p+\tilde{\sigma}^R_{q-p}}+\left[1+\mathcal{I}\left(\tfrac{p}{q}\right)\right]\frac{qp(p+q)}{\tilde{\sigma}^R_p+\tilde{\sigma}^R_{p+q}} \right\}.\nonumber
\end{eqnarray}
In the same way the kinetic equation can be modified to incorporate the vertex correction as well, and we find a set of coupled equations
\eq{DScEx}{
\left(\begin{array}{c}\partial_{\tau}n\\ \tilde{\sigma}^R\\ \mathcal{I} \end{array}\right)=\mathcal{F}(n,\tilde{\sigma}^R,\mathcal{I}).
}
They can be solved numerically according to the procedure described in Fig.~\ref{fig:Iteration}. Including the vertex correction, the second step of the iteration additionally includes the self-consistent determination of $\mathcal{I}$.
\begin{figure}
  \includegraphics[width=8.6cm]{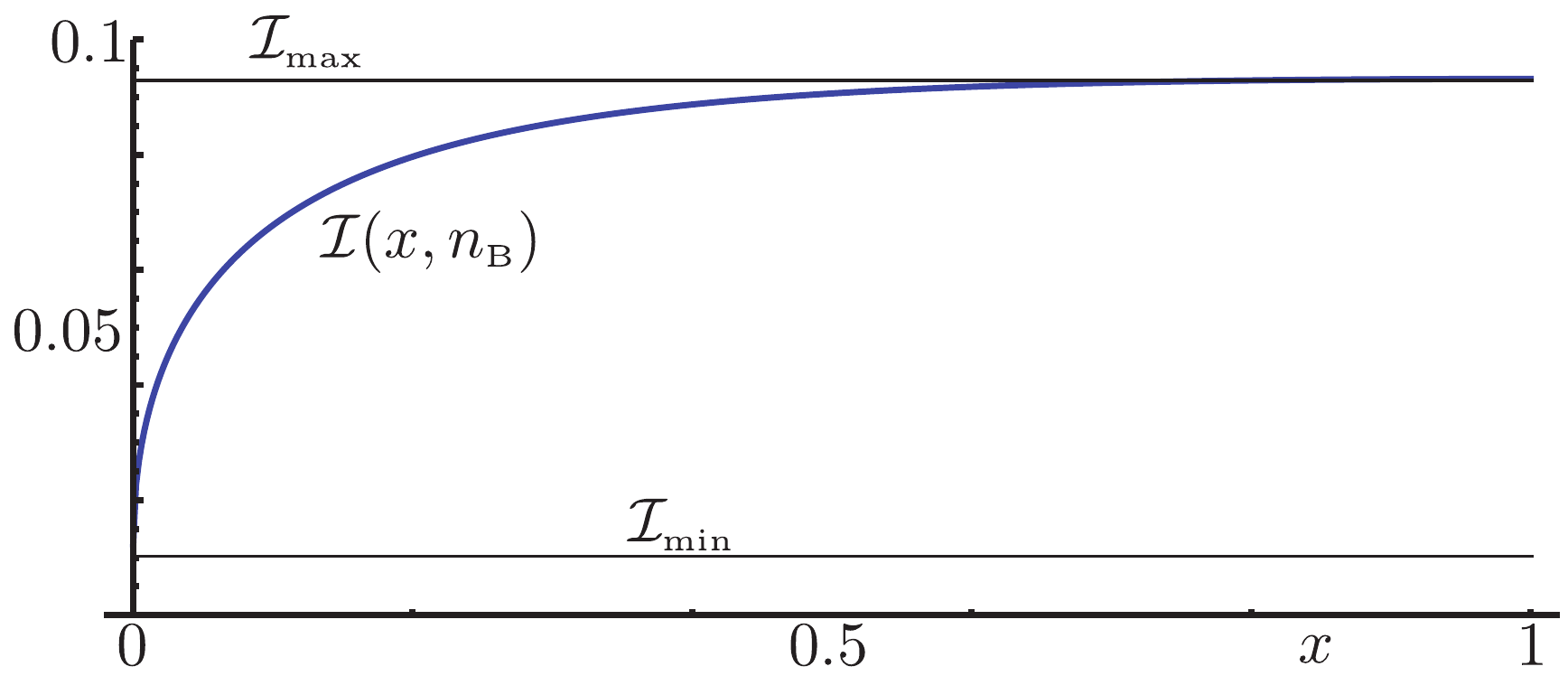}
  \caption{Vertex correction $\mathcal{I}(x,n\sub{B}(T))$ for an infinite temperature state. In the limit $T\rightarrow\infty$, the temperature drops out in Eq.~\eqref{DSc12}, and the vertex correction becomes temperature independent. Due to the invariance of $\mathcal{I}$ under $x\rightarrow \frac{1}{x}$, the plot is restricted to $0\leq x\leq 1$. The dependence of $\mathcal{I}$ on $x$ is weak, especially for $x\approx 1$, where it takes its maximum $\mathcal{I}(x,n\sub{B})\leq\mathcal{I}\sub{max}\approx0.093$.}
  \label{fig:VCorrPlot}
\end{figure}

We will now give an estimate of the order of the vertex correction for the case for which it is non-zero to estimate its impact on the dynamics and the kinetic equation. In the limit $T\rightarrow\infty$ the relevant phonon density is $n_q\approx \frac{T}{u|q|}$ and the self-energies have the thermal form $\tilde{\sigma}^R_q\propto\sqrt{\frac{T}{u}}q^{3/2}$. Consequently, the factor $\frac{T}{u}$ drops out and $\mathcal{I}(x,n)$ does no longer depend on temperature. In this case, $\mathcal{I}(0,n)\approx0.012, \mathcal{I}(1,n)\approx0.09$ and $\mathcal{I}(0,n)\leq \mathcal{I}(x,n)\leq\mathcal{I}(1,n)$. As can be seen in Fig.~\ref{fig:VCorrPlot}, the correction is small, with a weak momentum dependence. Consequently, the self-energy is only negligiably modified if instead of the full three-point vertex in Eq.~\eqref{DS9}, the bare value $\tilde{V}$ is used. This is precisely the self-consistent Born approximation that we used to determine the self-energies and the kinetic equation for the interacting Luttinger Liquid.

\section{Conclusion}
In this article, we used non-equilibrium field theory, in particular kinetic and Dyson-Schwinger equations, to determine the kinetics and non-equilibrium dynamics of resonantly interacting Luttinger Liquids. Exploiting the fact that the interactions lead to dressed but still well defined phonons, which enables a separation of timescales into slow forward and fast relative dynamics, we applied the Wigner and quasi-particle approximation and derived a closed set of simple yet powerful equations for the normal and anomalous phonon density, the phonon self-energy and vertex correction. These equations determine the dynamics of an interacting Luttinger Liquid initialized in a Gaussian (non-) equilibrium state. The resulting equations show strong aspects of universality, on the one hand being independent of any UV-scale, in particular independent of the Luttinger cutoff. On the other hand, after a proper rescaling of the forward time, all microscopic parameters entering the Hamiltonian can be eliminated and the only microscopic information entering the dynamical equations is the initial phonon density. We further used our approach to analytically determine the relaxation rate of a thermally excited state. For this dynamics, we found an initial exponential decay corresponding to previous results computed from linear response theory \cite{andreev80,zwerger06,samokhin98,affleck06,pereira08}. However, for longer times, the decay of the excitations follows a power law in time, revealing the presence of dynamical slow modes due to energy conservation. These latter modes are not contained in a plain equilibrium linear response theory but have to be build in by hand on the basis of symmetry arguments and conservation laws\cite{Lin13,Lux13}. Here, the dynamics based on the kinetic equation approach reveals the presence of these modes due to the algebraic decay at long times without any further modification, which shows the strength of our approach in a simple yet nontrivial example. 

The results of this work can be used in order to determine the kinetics and non-equilibrium dynamics of one-dimensional interacting quantum fluids prepared in a non-thermal initial state, which might occur as a consequence of a quantum quench or a sudden external perturbation. On the other hand, it paves the way to compute the dynamics of quantum fluids subject to drive and dissipation, and to determine the dynamics towards the steady state of an excited closed system. Of special interest and a strength of our approach is the treatment of long time dynamics in non-equilibrium systems, which are neither reachable by present numerical procedures nor by analytical approaches based on perturbation theory. Both examples belong to the uprising field of one-dimensional quantum fluids out-of-equilibrium and we leave their discussion open for future work.

\section{Acknowledgments}

We thank M. Heyl, S.~A.~E. Huber, A. Mitra for helpful discussions. This work was supported by the Austrian Science Fund (FWF) through SFB FOQUS F4016-N16 and the START grant Y 581-N16.

\bibliographystyle{abbrv}
\bibliography{Diss.bib}

\end{document}